\documentstyle[prd,aps,epsfig]{revtex}
\begin{document}

\rightline{}
\rightline{}

\Huge{\noindent{Istituto\\Nazionale\\Fisica\\Nucleare}}

\vspace{-3.9cm}

\Large{\rightline{Sezione SANIT\`{A}}}
\large{}
\rightline{Istituto Superiore di Sanit\`{a}}
\rightline{Viale Regina Elena 299}
\rightline{I-00161 Roma, Italy}

\vspace{0.95cm}

\rightline{INFN-ISS 97/15}
\rightline{November 1997}

\vspace{2cm}

\begin{center}

\Huge{\bf Rare exclusive semileptonic $b \to s$ transitions 
in the Standard Model}

\vspace{1.5cm}

\Large{D. Melikhov$^{a}, $ N. Nikitin$^{a}$ and S. Simula$^{b}$}

\vspace{1cm}

\large{$^{a}$Nuclear Physics Institute, Moscow State University\\
Moscow, 119899, Russia\\[0.15cm] $^{b}$Istituto Nazionale di Fisica 
Nucleare, Sezione Sanit\'a\\ Viale Regina Elena 299, I-00161 Roma, Italy}

\vspace{1cm}

\begin{abstract}
{\large 
We study long-distance effects in rare exclusive semileptonic decays 
$B \to (K, K^*) ~ (\ell^+ \ell^-, \nu \bar{\nu})$ and analyze dilepton 
spectra and asymmetries within the framework of the Standard Model. 
The form factors, describing the meson transition amplitudes of the 
effective Hamiltonian are calculated within the lattice-constrained 
dispersion quark model: the form factors are given by dispersion 
representations through the wave functions of the initial and final mesons,
and these wave functions are chosen such that the $B \to K^*$ transition form
factors agree with the lattice results at large $q^2$. We calculate branching 
ratios of semileptonic $B \to K, K^*$ transition modes and study the 
sensitivity of observables to the long-distance contributions. The shape 
of the forward-backward asymmetry and the longitudinal lepton polarization 
asymmetry are found to be independent of the long-distance effects and 
mainly determined by the values of the Wilson coefficients in the Standard 
Model.}
\end{abstract}

\end{center}

\newpage

\pagestyle{plain}

\normalsize{}

\title{Rare exclusive semileptonic $b \to s$ transitions 
in the Standard Model}

\author{D. Melikhov$^{a}, $ N. Nikitin$^{a}$ and S. Simula$^{b}$}
\address{$^{a}$Nuclear Physics Institute, Moscow State University,
Moscow, 119899, Russia\\ $^{b}$Istituto Nazionale di Fisica Nucleare,
Sezione Sanit\'a, Viale Regina Elena 299, I-00161 Roma, Italy}

\maketitle

\begin{abstract} 
We study long-distance effects in rare exclusive semileptonic decays 
$B\to(K,K^*)~(\ell^+\ell^-,\nu\bar{\nu})$ and analyze 
dilepton spectra and asymmetries within the framework of the Standard Model. 
The form factors, describing the meson transition amplitudes of the effective 
Hamiltonian are calculated within the lattice-constrained dispersion 
quark model: the form factors are given by dispersion representations 
through the wave functions of the initial and final mesons,
and these wave functions are chosen such that the $B\to K^*$ transition form
factors agree with the lattice results at large $q^2$. 
We calculate branching ratios of semileptonic $B\to K,K^*$ transition modes and 
study the sensitivity of observables to the
long-distance contributions. The shape of the forward-backward asymmetry and the 
longitudinal lepton polarization asymmetry are found
to be independent of the long-distance effects and mainly determined by the values of the 
Wilson coefficients in the Standard Model.
 
\end{abstract}

\section{Introduction}
The investigation of rare semileptonic decays of the $B$ meson induced 
by the flavour-changing neutral current transitions $b\to s$ represents 
an important test of the Standard Model (SM) and its possible extensions. 
Rare decays are forbidden at tree level and occur at the lowest order 
only through one-loop diagrams. This fact opens the
possibility to probe at comparatively low energies the structure of the
electroweak theory at large mass scales, thanks to the contributions of
virtual particles in the loops. Moreover, rare $b \to s$ transitions are
expected to be sensitive to possibile new interactions, like those
provided, e.g., by supersymmetric theories, two Higgs-doublet, top-color
and left-right models. These interactions govern the structure
of the operators and the corresponding Wilson coefficients, which appear
in the $\Delta B = 1$ effective electroweak Hamiltonian describing the
$b \to s$ transitions at low energies. 

A recent experimental observation of exclusive
\cite{cleo1} and inclusive \cite{cleo_incl} radiative decays, $B \to
K^* \gamma$ and $B \to X_s \gamma$, have prompted a lot of theoretical
investigation on rare semileptonic $B$ decays. However, in the 
case of exclusive decays any reliable extraction of the perturbative
(short-distance) effects encoded in the Wilson coefficients of the effective
 Hamiltonian 
\cite{gws,burasmuenz,ali,rescontr,nunu}
requires an accurate separation of the nonperturbative
(long-distance) contributions, which therefore should be known with
high accuracy. The theoretical investigation of these contributions
encounters the problem of describing the hadron structure, which 
provides the main uncertainty in the predictions of exclusive rare
decays. 

In exclusive $B\to K, K^*$ decays the long-distance effects in the meson 
transition amplitude of the effective Hamiltonian are encoded in the meson 
transition form factors of bilinear quark currents. 
Various theoretical frameworks have been
applied to the description of meson transition form factors; among them we 
should mention constituent quark models \cite{jauswyler,stech,gengkao,mns}, 
QCD sum rules \cite{colangelo,aliev,damir}, lattice QCD 
\cite{ape,ukqcd,lat}, 
approaches based on the heavy-quark symmetry \cite{burdman} and 
analytical constraints \cite{lellouch}. 

Lattice QCD simulations, because of its most direct connection with QCD, 
are expected to provide the most reliable results. 
Although it is not possible to place the $b$ quark directly on the lattice, a
constrained extrapolation in the heavy quark mass 
\cite{lat} allows to determine reliably the form factors for $B$ 
decays. A present limitation is that 
lattice calulations do not yet provide the form factors in the whole accessible
kinematical decay region: the daughter light quark 
produced in $b$ decay cannot move fast enough on the lattice
and one is therefore limited to the 
region of not very large recoils. For obtaining form factors in the whole
kinematical decay region one can 
use extrapolation procedures based on some parametrizations of the form
factors. For instance, in \cite{lat} a 
simple lattice-constrained parametrization based on the constituent quark
picture \cite{stech} and pole dominance is proposed. 
Anyway, a reliable knowledge of form factors in some region is already a
substantial step forward, which provides firm constraints for the results of 
other approaches. 

QCD sum rules give complementary information on the form factors 
as they can calculate the latter at not very large momentum 
transfers.  
However in practice various versions of the QCD sum 
rules give remarkably different predictions, being strongly
dependent on the technical subtleties of the particular version. 
A recent analysis \cite{ballbraun} disregards the three-point sum rules in
favour of the light-cone sum rules. 
On the other hand, the light-cone sum rules involve more phenomenological
inputs and the results 
turn out to be sensitive to the particular distribution amplitude of the light
meson used in the evaluation of the sum rule and to the model adopted for the 
subtraction of the continuum (cf. \cite{aliev} and \cite{damir}).  

Constituent quark models (QM) have proved to be a fruitful 
phenomenological method for the description of heavy meson transitions. 
An attractive feature of the approaches based on the concept of constituent
quarks is the suggestion of a simple physical picture of the decay process, 
based on the
following phenomena responsible for the soft physics: i) the chiral symmetry
breaking in the low-energy region generating the constituent quarks; ii) a 
strong 
peaking of the soft (nonperturbative) hadronic wave functions 
in terms of the relative constituent momenta with a width of the order of the 
confinement scale; and iii) the dominance of the contribution of the
Fock state components with the minimal number of constituents, i.e. 
$q\bar q$ component in mesons. An important shortcoming of the quark model 
predictions for the form factors is a strong dependence of the results on the 
QM parameters \cite{mns}. 

Thus we can see that none of the above-mentioned approaches can provide 
accurate form factors in the whole kinematically accessible region of $B$ 
decays. In this situation a combination of the results of  
different approaches can be efficient for obtaining reliable predictions. 

Our approach to the calculation of the $B\to K, K^*$ transitions 
is based on the dispersion quark model \cite{m1,m2}. The transition form 
factors are given by 
relativistic double spectral representations through the wave functions of 
the initial and final mesons both in the scattering and the decay regions. 
The form factors of the dispersion quark model develop 
the correct heavy-quark expansion at leading and next-to-leading $1/m_Q$ orders 
in accordance with QCD for the transitions between heavy quarks 
\cite{iwhh,luke}. For the heavy-to-light transition 
the form factors of the dispersion quark model satisfy the relations between 
the form factors 
of vector, axial-vector, and tensor currents valid at small recoil \cite{iwhl}. 
Thus the form factors of the dispersion quark model obey all known rigorous 
theoretical constraints. 
A possibility to calculate directly the form factors in all the decay region, 
avoiding in this way any extrapolation, is an important advantage of this 
formulation of the quark model. The main results of this work are as follows:
\begin{itemize}
\item We present a dispersion quark model calculation of the $B\to K, K^*$ 
transition form factors in the whole kinematical range of $q^2$. 
Adopting the quark masses and the wave functions of the Godfrey-Isgur  
model \cite{gi} for the hadron spectrum with a switched-off one-gluon exchange 
(OGE) potential for taking into account only the impact of the confinement 
scale, we have found that the resulting form factors are in good agreement 
with the lattice simulations at large $q^2$. Thus we expect to provide 
reliable form factors in the whole decay region. 

The dispersion quark model form factors for the $B\to K^*$ transition agree 
favorably in the 
whole range of $0<q^2<(M_B-M_K^*)^2$ with a lattice-constrained fit \cite{lat} 
based on the 
constituent quark picture \cite{stech} and an assumption on a single-pole 
behavior of the form factor $A_1(q^2)$. 
On the other hand the parametrizations based on heavy-quark symmetry (HQS) 
also give reasonable results if one assumes the leading-order expressions for 
the form factors and replaces the universal process-independent Isgur-Wise 
(IW) function with process-dependent form factors,
$\xi_{B\to K}$ and $\xi_{B\to K^*}$, related to the $B\to K$ and 
$B\to K^*$ transitions, respectively. 
The latter are found to differ strongly from each other and from the 
asymptotic IW function. 

An important consequence is that both our QM calculations of the form 
factors and the lattice-constrained parametrization of Ref. \cite{lat} 
as well as the use of the heavy-quark symmetry relations between 
the form factors predict a quite similar behavior for the forward-backward and 
lepton polarization asymmetries.  
\item We derive formulas for the differential decay rates and asymmetries in 
exclusive rare semileptonic decays of heavy mesons for the case of massive 
leptons and taking into account a nonzero mass of the daughter quark produced 
in the rare $b$ transition. For massless leptons and/or in the 
limit $m_s\to 0$ our formulas reproduce known results. 
\item We present a detailed analysis of non-resonant decay rates and 
asymmetries in $B \to(K,K^*)~(\ell^+\ell^-,\nu\bar{\nu})$ decays within the 
SM adopting our QM transition form factors. For comparison we also perform 
calculations with the lattice-constrained form factors of Ref. \cite{lat}.  
The decay rates evaluated in both models are found to be in agreement with 
each other, while the differential dilepton distributions are sensitive to 
subtle details of the $q^2$-dependence of the transition form factors. 
It is found that the lepton polarization asymmetry ($P_L$) as well as the 
shape of the forward-backward asymmetry $(A_{FB})$ are largely independent 
of the long-distance
contributions and determined only by the values of the Wilson coefficients 
in the SM. Such features make both $A_{FB}$ and $P_L$ good candidates for 
testing the Standard Model and probing possible New Physics.
\end{itemize}

The paper is organized as follows. In Section II the SM
operator basis, describing the $b \to s \ell^+ \ell^-$ and $b \to s \nu
\bar{\nu}$ transitions, is briefly presented. In Section III the meson
transition form factors are considered. Section IV presents the differential 
rates, lepton spectra and lepton asymmetries for the rare 
$B\to(K, K^*)~(\ell^+\ell^-,\nu\bar{\nu})$ decays including the 
case of massive leptons. Section V gives numerical analysis of the lepton spectra and lepton asymmetries 
in exclusive rare $B$-meson decays in the SM. 
Conclusion summarizes the results and gives an outlook.
 
\section{The operator basis}

The effective weak Hamiltonian, which
describes the $b \to s \ell^+ \ell^-$ transition, has the following form
\cite{gws}
\begin{equation}
\label{heff}
{\cal H}_{eff} = \frac{G_F}{\sqrt{2}} V_{tb} V_{ts}^\ast\, \sum_i
C_i(\mu) \, O_i(\mu),
\end{equation} 
where $G_F$ is the universal Fermi constant, the quantities $C_i(\mu)$ are
the Wilson coefficients, obtained after integrating out the heavy
particles, and the $O_i$'s are the basis operators; the sign of the Wilson coefficients 
$C_i(\mu)$ is determined as in the work
\cite{gws}: $C_2(M_W)=-1$. Within the SM, the
operators providing the main contribution to rare decays are
\cite{burasmuenz,ali}
\begin{eqnarray}
\label{basis}
O_1 &=& \left( \bar{s}_\alpha \gamma^\mu (1-\gamma_5)  b_\alpha \right) 
        \left( \bar{c}_\beta \gamma_\mu (1-\gamma_5) c_\beta \right), \nonumber \\
O_2 &=& \left( \bar{s}_\alpha \gamma^\mu (1-\gamma_5) b_\beta \right)  
        \left( \bar{c}_\beta \gamma_\mu (1-\gamma_5) c_\alpha \right), \nonumber \\
O_{7\gamma} &=& \frac{e}{8\pi^2}\bar{s}_\alpha \sigma_{\mu \nu}
            [m_b(\mu)(1+\gamma_5)+m_s(\mu)(1-\gamma_5)]b_\alpha\ F^{\mu \nu}, \nonumber \\
O_{9V} &=& \frac{e^2}{8\pi^2}(\bar{s}_\alpha \gamma^\mu(1-\gamma_5) b_\alpha) \bar{l}
         \gamma_\mu l,  \nonumber \\
O_{10A} &=& \frac{e^2}{8\pi^2}(\bar{s}_\alpha \gamma^\mu(1-\gamma_5) b_\alpha)
         \bar{l} \gamma_\mu \gamma_5 l, \nonumber \\
\end{eqnarray} 
In Eq. (\ref{heff}) the renormalization scale $\mu$ is usually
chosen to be $\mu \simeq m_b$ in order to avoid large logarithms in the
matrix elements of the operators $O_i$. The Wilson coefficients $C_i$
reflect the specific features of the theory at large mass scales; they
are calculated at the scale $\mu \simeq M_W$ and then evolved down to
$\mu\simeq m_b$ by the renormalization group equations. The analytic
expressions for $C_i(\mu)$ in the SM can be found, e.g., in
\cite{burasmuenz}. In what follows, the values of the Wilson coefficients
at the scale $\mu\simeq m_b\simeq 5\;GeV$ are \cite{gws,burasmuenz}: 
$C_1(m_b) = 0.241$,
$C_2(m_b) = -1.1$, 
$C_{7\gamma}(m_b) = 0.312$, 
$C_{9V}(m_b) =-4.21$ and 
$C_{10A}(m_b)= 4.64$.

The four-quark operators $O_1$ and $O_2$ generate both short- and
long-distance contributions to the effective weak Hamiltonian
(\ref{heff}). Both contributions can be taken into account by replacing
$C_{9V}(m_b)$ with an effective coefficient $C_{9V}^{eff}(m_b, q^2)$ given by
\cite{ali}
\begin{equation} 
\label{c9eff}
C_{9V}^{eff}(m_b,q^2)=C_{9V}(m_b)+\left[3C_1(m_b)+C_2(m_b)\right]\cdot
\left[h\left({m_c \over m_b}, {q^2 \over m_b^2}\right) +
\frac{3}{\alpha_{em}^2}\kappa\sum_{V_i=J/\psi,\psi',\cdots}
\frac{\pi \Gamma(V_i \to \ell \ell) M_{V_i}}{M_{V_i}^2 - q^2 - i
M_{V_i} \Gamma_{V_i}} \right], 
\end{equation}  
where $q^2$ is the invariant mass squared of the lepton pair. 
The short-distance contributions are contained in the function $h(m_c/m_b,
q^2/m_b^2)$, which describes the one-loop matrix element of the four-quark
operators $O_1$ and $O_2$ (see, e.g., \cite{burasmuenz} for its explicit
expression). The long-distance contribution, related to the formation of
intermediate $c \bar{c}$ bound states, is usually estimated by combining
the factorization hypothesis and the Vector Meson Dominance 
assumption \cite{ali,rescontr}; phenomenological analyses \cite{rescontr}
suggest that in order to reproduce correctly the branching ratio
${\rm BR}(B \to J / \psi X \to \ell^+ \ell^- X) =$ ${\rm BR}(B \to J /
\psi X) ~ \cdot$ ${\rm BR}(J / \psi \to \ell^+ \ell^-)$ the fudge factor
$\kappa$, which appears in Eq. (\ref{c9eff}) to correct
phenomenologically for inadequacies of the factorization + VMD
framework, should satisfy the approximate relation: 
$\kappa ~ \left[ 3
C_1(m_b) + C_2(m_b) \right] \approx 1$. To sum up, the effective weak
Hamiltonian has the following structure (cf. \cite{burasmuenz,ali,nunu})
\begin{eqnarray}
\label{heffbtosll}
{\cal H}_{eff}(b\to sl^{+}l^{-})&=&{\frac{G_{F}}{\sqrt2}}{\frac{\alpha_{em}}{2\pi}}
V^{*}_{ts}\,V_{tb}                                            
\left[\,-2{\frac{C_{7\gamma}(m_b)}{q^2}}
\left((m_b+m_s)({\bar s}i\sigma_{\mu\nu}q^{\nu}b)+
(m_b-m_s)({\bar s}i\sigma_{\mu\nu}q^{\nu}\gamma_{5}b) \right)
({\bar l}\gamma^{\mu}l)\right.\\
&&+\left. C_{9V}^{eff}(m_b, q^2)({\bar s}\gamma_{\mu}
(1-\gamma_{5})b)({\bar l}\gamma^{\mu}l)+
C_{10A}(m_b)({\bar s}\gamma_{\mu}(1-\gamma_{5})b)
({\bar l}\gamma^{\mu}\gamma_{5}l)\right]  \nonumber
\end{eqnarray}
\begin{equation}
\label{Hneutrino}
{\cal H}_{eff}(b\to s\nu\hat\nu)=
\frac{G_{F}}{\sqrt{2}}\frac{\alpha_{em}}{2\pi \sin^2{\theta_W}}
V_{tb}V^{*}_{ts}X(x_t)
({\bar s}_{\alpha}\gamma_{\mu}(1-\gamma_{5})b^{\alpha})({\bar
\nu}\gamma^{\mu}(1-\gamma_{5})\nu)
\end{equation}
where $x_t = (m_t/M_W)^2$ and $X(x_t)$ is given in \cite{nunu}. At $m_t =
176 ~ GeV$ one has $X(x_t) = 2.02$. 
\newpage

\section{Meson transition form factors.} The long-distance contribution to $B
\to (K, K^*)$ decays is contained in the meson matrix elements of the
bilinear quark currents of ${\cal{H}}_{eff}$, i.e. in the
relativistic invariant transition form factors of the vector,
axial-vector and tensor currents. In rare semileptonic decays
there is another long-distance effect, known as the weak annihilation,
which is caused by the Cabibbo-suppressed part of the four-fermion
operators not included in the operator basis (\ref{heff}). However, the
impact of this process in $B \to (K, K^*)$ transitions is negligible
\cite{ali}. The amplitudes of meson decays are induced by the quark
transition $b \to s$ through the vector $V_{\mu} = \bar{s} \gamma_{\mu}
b$, axial-vector  $A_{\mu} = \bar{s} \gamma_{\mu} \gamma^5 b$,  tensor
$T_{\mu\nu} = \bar{s} \sigma_{\mu\nu} b$, and pseudotensor 
$T^5_{\mu\nu}\;=\;{\bar q}\sigma_{\mu\nu}\gamma_5b$ currents, 
with the following covariant structure \cite{iwhl}
\begin{eqnarray}
\label{amplitudes}
<P(M_2,p_2)|V_\mu(0)|P(M_1,p_1)>&=&f_+(q^2)P_{\mu}+f_-(q^2)q_{\mu},  \nonumber \\
<V(M_2,p_2,\epsilon)|V_\mu(0)|P(M_1,p_1)>&=&2g(q^2)\epsilon_{\mu\nu\alpha\beta}
\epsilon^{*\nu}\,p_1^{\alpha}\,p_2^{\beta}, \nonumber \\
<V(M_2,p_2,\epsilon)|A_\mu(0)|P(M_1,p_1)>&=&
i\epsilon^{*\alpha}\,[\,f(q^2)g_{\mu\alpha}+a_+(q^2)p_{1\alpha}P_{\mu}+
a_-(q^2)p_{1\alpha}q_{\mu}\,],   \nonumber \\
<P(M_2,p_2)|T_{\mu\nu}(0)|P(M_1,p_1)>&=&-2i\,s(q^2)\,(p_{1\mu}p_{2\nu}-p_{1\nu}p_{2\mu}), \nonumber
\\
<V(M_2,p_2,\epsilon)|T_{\mu\nu}(0)|P(M_1,p_1)>&=&i\epsilon^{*\alpha}\,
[\,g_{+}(q^2)\epsilon_{\mu\nu\alpha\beta}P^{\beta}+
g_{-}(q^2)\epsilon_{\mu\nu\alpha\beta}q^{\beta}+
g_0(q^2)p_{1\alpha}\epsilon_{\mu\nu\beta\gamma}p_1^{\beta}p_2^{\gamma}\,], \nonumber \\ 
<P(M_2,p_2)|T^5_{\mu\nu}(0)|P(M_1,p_1)>&=&s(q^2)\,\epsilon_{\mu\nu\alpha\beta}P^\alpha q^\beta
,\nonumber \\
<V(M_2,p_2,\epsilon)|T^5_{\mu\nu}(0)|P(M_1,p_1)>&=&
g_+(q^2)(\epsilon_\nu^*P_\mu-\epsilon_\mu^*P_\nu) 
+g_-(q^2)(\epsilon_\nu^*q_\mu-\epsilon_\mu^*q_\nu) \nonumber \\
&&+g_0(q^2)(p_1\epsilon^*)(p_{1\nu}p_{2\mu}-p_{1\mu}p_{2\nu})
\end{eqnarray}
where $q=p_{1}-p_{2}$, $P=p_{1}+p_{2}$. We use the following notations: 
$\gamma^{5}=i\gamma^{0}\gamma^{1}\gamma^{2}\gamma^{3}$, 
$\sigma_{\mu \nu}={\frac{i}{2}}[\gamma_{\mu},\gamma_{\nu}]$,
$\epsilon^{0123}=-1$,  
$\gamma_{5}\sigma_{\mu\nu}=-{\frac{i}{2}}\epsilon_{\mu\nu\alpha\beta}\sigma^{\alpha\beta}$,
and 
$Sp(\gamma^{5}\gamma^{\mu}\gamma^{\nu}\gamma^{\alpha}\gamma^{\beta})=
4i\epsilon^{\mu\nu\alpha\beta}$.

Another frequently used set of the form factors is connected with the set (\ref{amplitudes})
as follows 
\begin{eqnarray}
\label{alt_ff}
F_1(q^2) & = & f_+(q^2), \nonumber \\
F_0(q^2) & = & f_+(q^2) + q^2 f_-(q^2) / (Pq), \nonumber \\
F_T(q^2) & = & -(M_1 + M_2) s(q^2), \nonumber \\
V(q^2) & = & (M_1 + M_2) g(q^2),  \nonumber \\
A_1(q^2) & = & f(q^2) / (M_1 + M_2),  \nonumber \\
A_2(q^2) & = & -(M_1 + M_2) a_+(q^2),  \nonumber \\
A_0(q^2) & = & [q^2 a_-(q^2) + f(q^2) + (Pq) a_+(q^2)] / 2M_2, 
\nonumber \\
T_1(q^2) & = & -g_+(q^2)/2,  \nonumber \\
T_2(q^2) & = &-(g_{+}(q^2)+q^2g_{-}(q^2)/(Pq))/2,  \nonumber \\
T_3(q^2) & = & (M_1 + M_2)^2 [g_-(q^2) / (Pq) - h(q^2) / 2]/2
\end{eqnarray}

The relativistic invariant form factors encode the dynamical information about the decay process and 
should be considered within a nonperturbative approach. 
We investigate the meson form factors within a dispersion formulation of the 
relativistic constituent quark model (QM) model
\cite{m1}.  

Let us consider the transition from the initial meson $q(m_2)
\bar{q}(m_3)$ with mass $M_1$ to the final meson $q(m_1) \bar q(m_3)$
with mass $M_2$, induced by the quark transition $m_2 \to m_1$ through
the current $\bar{q}(m_1) J_{\mu (\nu)} q(m_2)$. For the transition $B_u
\to (K, K^*)$ one has $m_2 = m_b$, $m_1 = m_s$ and $m_3 = m_u$. The
$CQ$ structure of the initial and final mesons is described by the
vertices $\Gamma_1$ and $\Gamma_2$, respectively. The initial $B$-meson
vertex has the spinorial structure $\Gamma_1 = i\gamma_5 ~ G_1 /
\sqrt{N_c}$, where $N_c$ is the number of colours; the final meson vertex
has the structure $\Gamma_2 = i\gamma_5 ~ G_2 / \sqrt{N_c}$ for a
pseudoscalar state and $\Gamma_{2\mu} = [A \gamma_\mu + B(k_1 - k_3)_\mu]
~ G_2 / \sqrt{N_c}$, with $A = -1$ and $B = 1 / (\sqrt{s_2} + m_1 + m_3)$
for an $S$-wave vector meson. 
At $q^2<0$ the spectral representations of the form factors have the form \cite{m1}
\begin{equation}
\label{ff}
f_i(q^2)=
\frac1{16\pi^2}\int\limits^\infty_{(m_1+m_3)^2}ds_2\varphi_2(s_2)
\int\limits^{s_1^{+}(s_2,q^2)}_{s_1^{-}(s_2,q^2)}ds_1\varphi_1(s_1)
\frac{\tilde f_i(s_1,s_2,q^2)}{\lambda^{1/2}(s_1,s_2,q^2)},
\end{equation}
where the wave function $\varphi_i(s_i)=G_i(s_i)/(s_i-M_i^2)$ and 
$$
s_1^\pm(s_2,q^2)=
\frac{s_2(m_1^2+m_2^2-q^2)+q^2(m_1^2+m_3^2)-(m_1^2-m_2^2)(m_1^2-m_3^2)}{2m_1^2}
\pm\frac{\lambda^{1/2}(s_2,m_3^2,m_1^2)\lambda^{1/2}(q^2,m_1^2,m_2^2)}{2m_1^2}
$$
and 
$
\lambda(s_1,s_2,s_3)=(s_1+s_2-s_3)^2-4s_1s_2
$
is the triangle function. 

The Eq. (\ref{ff}) accounts for only two-particle singularities in the Feynman graphs. For
ground-state pseudoscalar and vector mesons built up of constituent quarks with
the masses $m_q$ and $m_{\bar{q}}$, the function $\varphi(s)$ can be written as 
\begin{equation}
\label{vertex}
\varphi(s) = \frac{\pi}{\sqrt{2}} \frac{\sqrt{s^2 - (m_q^2 -
m_{\bar{q}}^2)^2}} {\sqrt{s - (m_q - m_{\bar{q}})^2}} \frac{w(k^2)}{s^{3/4}}, 
\end{equation}
where $k = \lambda^{1/2}(s, m_q^2, m_{\bar{q}}^2) / 2 \sqrt{s}$ and
$w(k^2)$ is the ground-state $S$-wave radial wave function, normalized as
$\int_0^{\infty} dk k^2 |w(k^2)|^2 = 1$. 

The unsubtracted double spectral densities $\tilde f_i(s_1,s_2,q^2)$ of the form factors
read \cite{m2}: 
\begin{eqnarray}
\label{s}
\tilde s&=&2\,[m_1\alpha_{2}+m_2\alpha_{1}+m_3(1-\alpha_{1}-\alpha_{2})],
\\
\label{f1}
\tilde f_++\tilde f_-&\equiv&\tilde f_1=2m_1\tilde s+4\alpha_2[s_2-(m_1-m_3)^2]-2m_3\tilde s, 
\\
\label{f2}
\tilde f_+-\tilde f_-&\equiv&\tilde f_2=2m_2\tilde s+4\alpha_1[s_1-(m_2-m_3)^2]-2m_3\tilde s,
\\
\label{g}
\tilde g&=&-A\tilde s-4B\beta,
\\
\label{g1}
\tilde g_++\tilde g_-&=&A\tilde f_1-8\beta+8B(m_1+m_3)\beta,
\\
\label{g2}
\tilde g_+-\tilde g_-&=&A\tilde f_2+8B(m_2-m_3)\beta,
\\
\label{a2}
\tilde a_{+D}-\tilde a_{-D}&=&-2\tilde s+ 4BC_2\alpha_1+\alpha_{12}C_0,
\\
\label{a1}
\tilde a_{+D}+\tilde a_{-D}&=&-4A\,(2m_3+BC_1)\alpha_1+\alpha_{11}C_0,
\\
\label{f}
\tilde f_D&=&-4A[m_1m_2m_3+\frac{m_2}2(s_2-m_1^2-m_3^2)
+\frac{m_1}2(s_1-m_2^2-m_3^2)-\frac{m_3}2(s_3-m_1^2-m_2^2)]+C_0\beta,
\\
\label{g0}
\tilde
g_{0D}&=&-8A\alpha_{12}-8B\,[-m_3\alpha_{1}+(m_3-m_2)\alpha_{11}+(m_3+m_1)\alpha_{12}],
\end{eqnarray}
where 
\begin{eqnarray}
\label{alpha1}
\alpha_1&=&\left[(s_1+s_2-s_3)(s_2-m_1^2+m_3^2)-2s_2(s_1-m_2^2+m_3^2)\right]/{\lambda(s_1,s_2,s_3)},
\\
\label{alpha2}
\alpha_2&=&
\left[(s_1+s_2-s_3)(s_1-m_2^2+m_3^2)-2s_1(s_2-m_1^2+m_3^2)\right]/{\lambda(s_1,s_2,s_3)},
\\
\label{beta}
\beta&=&\frac14\left[2m_3^2-\alpha_1(s_1-m_2^2+m_3^2)-\alpha_2(s_2-m_1^2+m_3^2)\right],
\\
\label{alpha11}
\alpha_{11}&=&\alpha_1^2+4\beta {s_2}/{\lambda(s_1,s_2,s_3)}, \quad 
\alpha_{12}=\alpha_1\alpha_2-2\beta(s_1+s_2-s_3)/\lambda(s_1,s_2,s_3),
\\
\label{cc}
C_0&=&-8A(m_2-m_3)+4BC_3,\quad C_1=s_2-(m_1+m_3)^2, 
\\
C_2&=&s_1-(m_2-m_3)^2, \quad C_3=s_3-(m_1+m_2)^2-C_1-C_2. \nonumber
\end{eqnarray}

We label with a subscript 'D' the double spectral densities of the form factors which 
require subtractions. 
The subtraction procedure has been fixed by matching the $1/m_Q$ expansion of the form 
factors in the quark model to the corresponding expansion in QCD 
in leading and next--to--leading orders for the case 
of a meson transition caused by heavy--to--heavy quark
transition \cite{m2}. 

The double spectral densities with properly defined subtraction terms read 
\begin{eqnarray}
\label{fsub}
\tilde f&=&\tilde f_D+[(M_1^2-s_1)+(M_2^2-s_2)]\tilde g, \\
\label{a_+sub}
\tilde a_+&=&\tilde a_{+D}
+\frac{\sqrt{s_1}+\sqrt{s_2}}{(\sqrt{s_1}+\sqrt{s_2})^2-s_3}
\left(\frac{M_1^2-s_1}{\sqrt{s_1}}+\frac{M_2^2-s_2}{\sqrt{s_2}}\right)
\frac{\tilde g}2,\\
\label{a2+sub}
\tilde a_-&=&\tilde a_{-D}
+\frac{\sqrt{s_2}-\sqrt{s_1}}{(\sqrt{s_1}+\sqrt{s_2})^2-s_3}
\left(\frac{M_1^2-s_1}{\sqrt{s_1}}+\frac{M_2^2-s_2}{\sqrt{s_2}}\right)
\frac{\tilde g}2,\\
\label{g0+sub}
\tilde g_0&=&\tilde g_{0D}
+\frac{1}{(\sqrt{s_1}+\sqrt{s_2})^2-s_3}\left(\frac{M_1^2-s_1}{\sqrt{s_1}}+\frac{M_2^2-s_2}{\sqrt{s_2}}\right)
\tilde g.
\end{eqnarray}

The structure of the HQ expansion in LO and NLO $1/m_Q$ orders 
of the form factors given by the Eq. (\ref{ff}) with the spectral densities
(\ref{s}-ref{g0}) 
agrees with the corresponding structure of the HQ expansion in QCD 
\cite{luke}, provided that the functions
$\varphi(s_i)$ are localized near the $q \bar{q}$ threshold with a width of
the order $\Lambda_{QCD}$ \cite{m2}. Moreover, for the case of meson decays induced
by a heavy-to-light quark transition the dispersion formulation provides the form facotrs 
which satisfy the leading-order relations between the form factors of the vector and
tensor currents near zero recoil given in \cite{iwhl}. 

As the analytical continuation to the time-like region $q^2 > 0$ is performed, in addition to the 
normal contribution which is just the expression (\ref{ff}) taken at $q^2 > 0$, 
the anomalous contribution, described explicitly
in \cite{m1} emerges. The normal contribution dominates the form factors at small
$q^2$ and vanishes when $q^2 = (m_2 - m_1)^2$, while the anomalous
contribution is negligible at small $q^2>0$ and steeply rises as $q^2 \to
(m_2 - m_1)^2$. 

Notice that since the dispersion quark model is based on taking into account only two-particle $q\bar q$ 
intermediate states in the amplitude of the interaction of the $q\bar q$ constituent quark pair with
the external field it is conceptually close to the light-cone quark model LCQM \cite{jauswyler}. 
In particular, the form factors of the LCQM \cite{jauswyler} can be rewritten at $q^2<0$ as double spectral 
representations similar to the dispersion model. One finds that at $q^2<0$ the form 
factors which are given by the unsubtracted spectral representations in the dispersion formulation are the same as 
in the LCQM. At the same time, the LCQM form factors $f$, $a_1$, $a_2$, and $h$ are different 
from the dispersion quark model form factors and do not develop a correct heavy quark expansion in the 
next-to-leading $1/m_Q$ order. 

For evaluating the form factors we need to specify the quark model parameters such as the
constituent quark masses and the wave functions. 
In Ref. \cite{mns} we have run calculations of the mesonic form
factors (\ref{ff}) adopting different QMs for the radial
wave function $w(k^2)$ appearing in Eq. (\ref{vertex}), in particular:
a simple Gaussian ansatz of the ISGW2 model \cite{isgw2} and the
variational solution \cite{sim95} of the effective $q \bar{q}$
semi-relativistic Hamiltonian of Godfrey and Isgur (GI) \cite{gi}.
These two models differ both in the shape of the radial wave function,
particularly at high momenta $k$, and in the values of the quark masses (see Ref.\cite{mns}). 
The results of our calculations have
shown that the mesonic form factors (\ref{ff}) are sensitive both to the
high-momentum tail of the meson wave function and to the values adopted
for the quark masses (see also Refs. \cite{gns96,sim96}). 

In order to obtain more reliable predictions for the form factors, we require the QM parameters to be adjusted in
such a way that the calculated form factors at large $q^2$ are compatible with the lattice results 
\cite{ape,ukqcd}. We have found that the best agreement with the lattice data at large $q^2$ is obtained for the 
quark masses and wave functions of the GI model with a switched-off one-gluon exchange (GI-OGE). 
The constituent quark masses and the average momenta squared characterizing the GI-OGE model are given in Table 
\ref{table:parameters}. Table \ref{table:fits} presents simple fits to the calculated form factors. 

\begin{center}
\begin{figure}
\begin{tabular}{cc}
\mbox{\epsfig{file=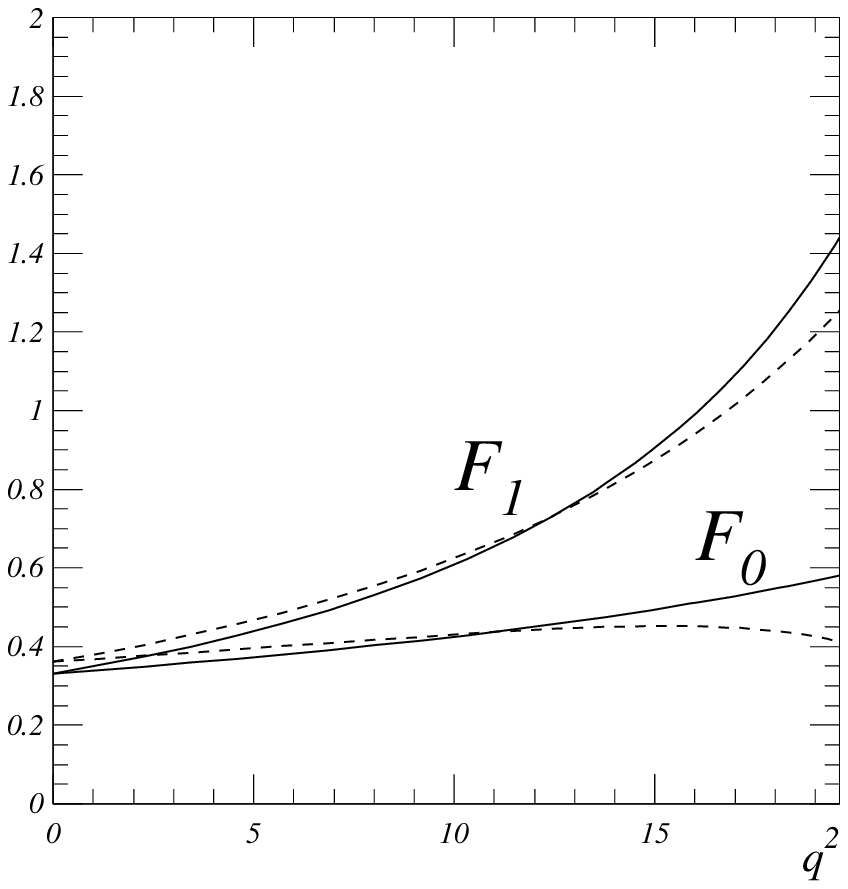,width=7.cm}}
& 
\mbox{\epsfig{file=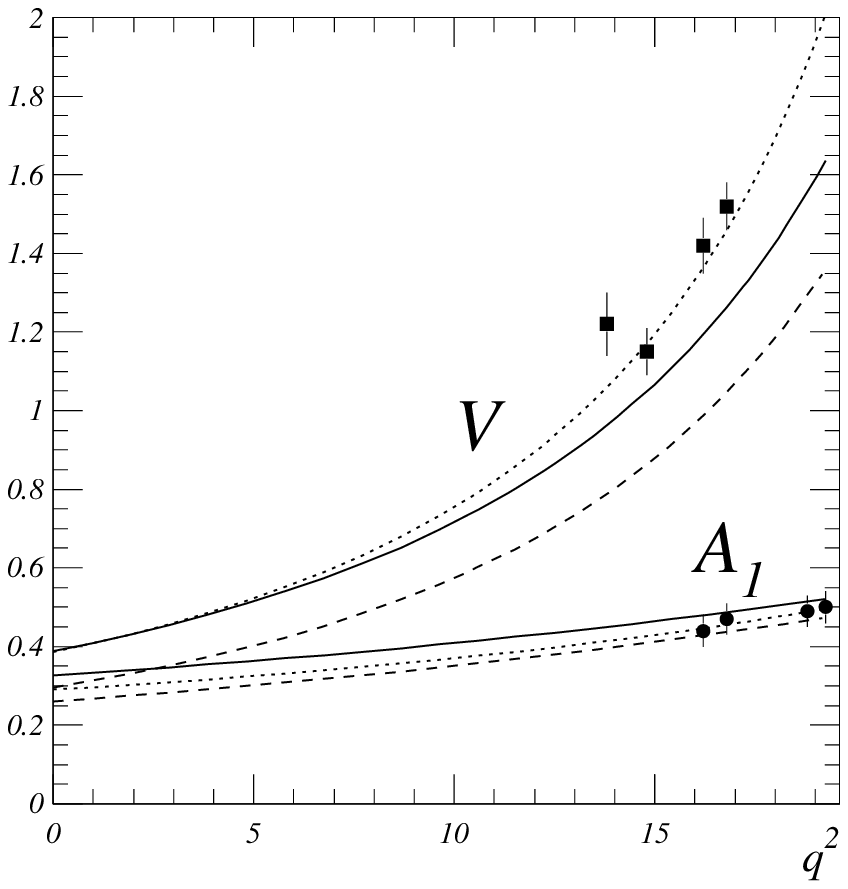,width=7.cm}}\\
\mbox{\epsfig{file=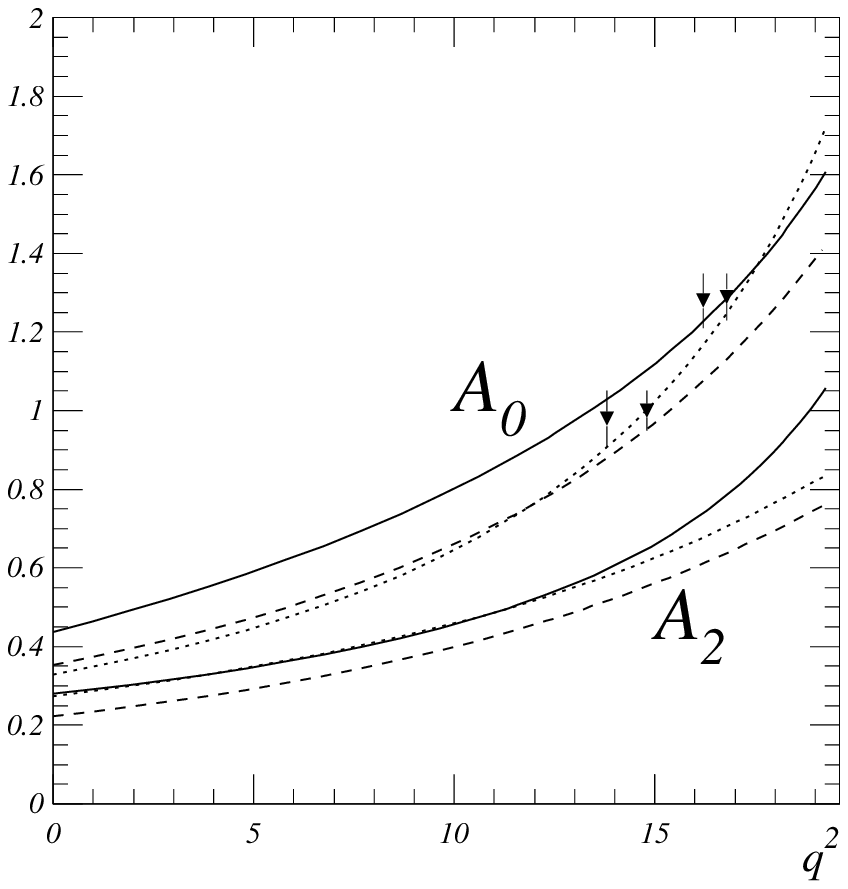,width=7.cm}}
& 
\mbox{\epsfig{file=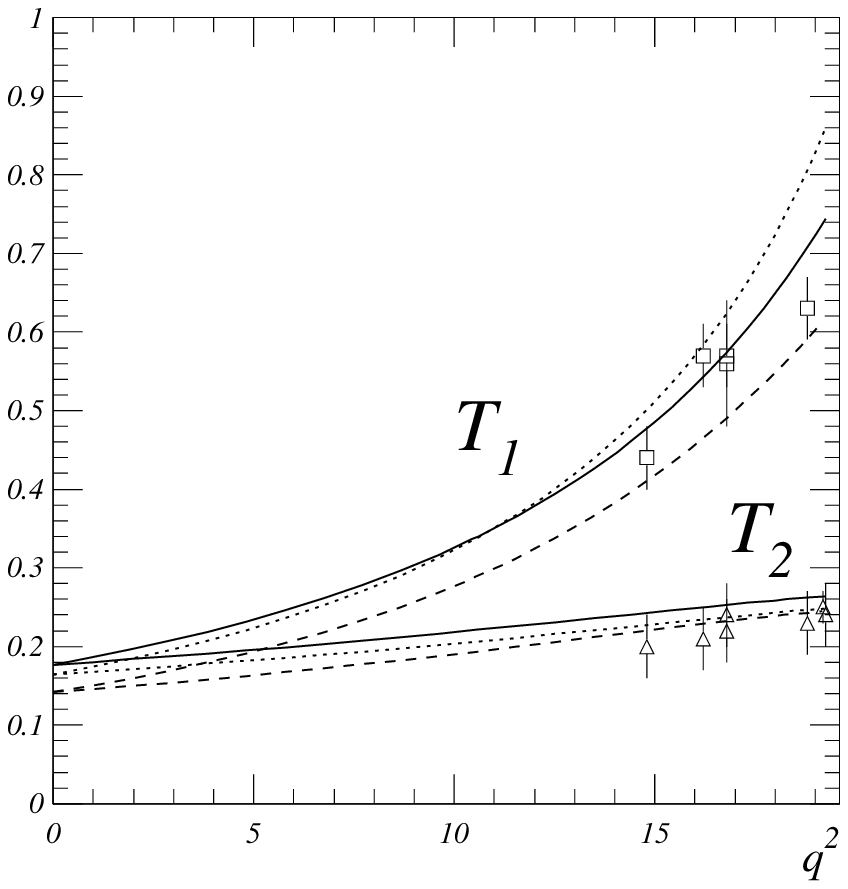,width=7.cm}}
\end{tabular}
\caption{\label{fig:ffacts}
Form factors for the $B\to K,K^*$ transition vs the lattice data: solid - GI-OGE, dashed - ISGW2 models, dotted - 
lattice-constrained parametrization of \protect\cite{lat}.}
\end{figure}
\end{center}

Fig. \ref{fig:ffacts} shows the GI-OGE form factors versus the available lattice data for $B\to K^*$ 
\cite{ape,ukqcd}. For comparison, we also present the form factors obtained within the ISGW2 exponential ansatz 
for the soft meson wave functions \cite{mns}. 

A good agreement of our QM predictions obtained with the GI-OGE wave functions with the results of lattice 
simulations is not surprising: the strong long-distance physics is dominated by the confinement mechanism, and
therefore it seems quite natural, that soft wave functions which take into account the effects of the confinement
scale provide the form factors in agreement with the lattice QCD results.

\begin{table}[hbt]
\caption{\label{table:parameters}
Constituent quark masses (in $GeV$) and the average momentum squared (in $GeV^2$).} 
\centering
\begin{tabular}{|c||c|c|c|c||c|c|c|} 
Ref. & $m_u$ & $m_s$ & $m_c$ & $m_b$ & $<k^2>_K$ & $<k^2>_{K^*}$ & $<k^2>_{B_u}$\\ 
\hline 
GI-OGE \cite{sim95} & 0.22 & 0.42 & 1.65 & 5.0 & 0.17 & 0.17  & 0.26   
\end{tabular}
\end{table}

\begin{table}[b]
\caption{\label{table:fits}Parameters of the fit $f_i(q^2) = f_i(0) / [1 -
\sigma_1 q^2 + \sigma_2 q^4]$ to the $B \to (K, K^*)$ transition form
factors in the GI-OGE model.} 
\centering
\begin{tabular}{|c||c|c|c||c|c|c|c|c|c|c|}
Decay&\multicolumn{3}{c|}{$B \to K$} & \multicolumn{7}{c|}{$B \to K^*$}\\ 
\hline
     &$f_{+}(0)$ & $f_{-}(0)$ & $s(0)$ & $g(0)$   & $f(0)$ & $a_+(0)$   & $a_{-}(0)$ & $h(0)$ & $g_+(0)$ & $g_-(0)$\\  
Ref. & $\sigma_1$&$\sigma_1$&$\sigma_1$&$\sigma_1$&$\sigma_1$ &$\sigma_1$&$\sigma_1$&$\sigma_1$&$\sigma_1$&$\sigma_1$\\
     & $\sigma_2$&$\sigma_2$&$\sigma_2$&$\sigma_2$&$\sigma_2$ &$\sigma_2$&$\sigma_2$&$\sigma_2$&$\sigma_2$&$\sigma_2$\\
\hline 
 GI-OGE   &  0.33    &$-$0.27    & 0.057   & 0.063   & 2.01   &$-$0.0454  &  0.053  & 0.0056  &$-$0.3540 & 0.313\\
          &  0.0519  &   0.0524  & 0.0517  & 0.0523  & 0.0212 &   0.039   &  0.044  & 0.0657  &   0.0523 & 0.053\\
          &  0.00065 &   0.00066 & 0.00064 & 0.00066 & 0.00009&   0.00004 &  0.00023& 0.0010  &   0.0007 & 0.00067 
\end{tabular}
\end{table}

Recently, a lattice-constrained parametrization for the $B\to K^*$ form factors has been developed in 
the whole range of accessible values of $q^2$ \cite{ukqcd}. 
It is based on the Stech parametrization of the form factors obtained within the constituent 
quark picture, HQS scaling relations near $q^2 = q_{max}^2$ and a single-pole behavior of $A_1$
suggested by $m_Q$-scaling relations at $q^2=0$ from the LCSRs in the HQ limit.  
Parameters of the single-pole fit to the form factor $A_1(0)=0.29^{+0.04}_{-0.03}$ and
$M_1=6.8^{+0.7}_{-0.4}\;GeV$ are found from the least-$\chi^2$ fit 
to the lattice QCD simulation in a limited region at high values of $q^2$. 
Such a parametrization, though still phenomenological, is also consistent with the dispersive bounds of Ref.
\cite{lellouch} and therefore obeys all known theoretical constraints. 
It should be taken into account however that the lattice-constrained parametrization is an approximation:  
in particular, it suggests the relation $T_1(q^2)=(1-q^2/Pq)T_2(q^2)$ which can be also 
translated into $g_+(q^2)=-g_-(q^2)$. In dynamical calculation within QMs or LCSRs these 
relations are fulfilled within 10\% accuracy but are never found to be exact. 
Nevertheless, approximate Stech's relations combined with a monopole fit to $A_1$ exhibit surprisingly good agreement 
with the lattice points at large $q^2$ (see Fig. \ref{fig:ffacts}). 

The form factor $A_1$ calculated in our approach for the GI-OGE model wave functions is found to have a 
behavior very close to the single-pole function with the parameters $A_1(0)=0.326$ and $M_1=6.86\; GeV$ 
in agreement with an assumption of Ref. \cite{lat}. The results on most of the form factors are within 5\% 
agreement with the parametrizations \cite{lat} except for the form factor $V$ which turns out to be at zero recoil 
some 15\% smaller in our calculations. 

Table \ref{table:comparison} compares predictions on the form factors from various approaches. 
One can see that our results agree with those of the LCSR of Ref. \cite{damir}. 
The form factors of another version of the LCSR \cite{aliev} have different behavior which disagrees 
also with the lattice results at large $q^2$: namely, at $q^2\simeq 16\;GeV^2$ the 
form factors $T_1,T_2$, and $A_1$ turn out to be considerably larger than the lattice points, and  
the form factor $A_0$ to be too small. The form factors $T_2$ and $A_1$ in the 3ptSR approach \cite{colangelo} 
are decreasing with $q^2$ in contradiction with the results of other approaches and lattice simulations. 

\begin{figure}
\centering
\mbox{\epsfig{file=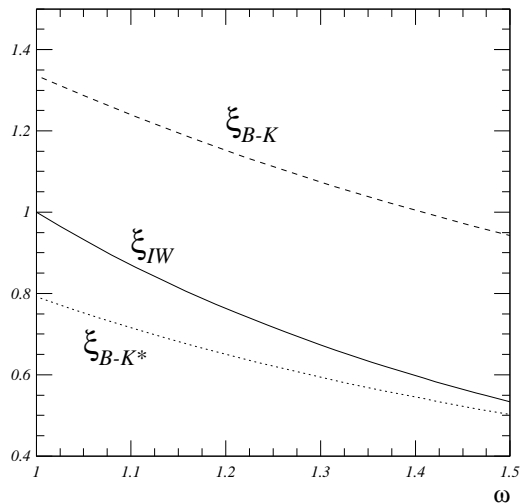,width=7.cm}}
\caption{\label{fig:xi}
The Isgur-Wise function $\xi$ (solid) \protect\cite{m2,sim96}, and the form 
factor $\xi_{B\to K}$ (dotted) and $\xi_{B\to K^*}$ (dashed) calculated via 
Eq. (\protect\ref{xi}). }
\end{figure}

Let us notice, that an approximate relation $g_+=-g_-$ found to describe well the lattice points 
at large $q^2$ and extended in \cite{lat} to the whole kinematically accessible region, might signal 
that relations motivated by the heavy-quark symmetry also work with a reasonable accuracy in the $B\to K,K^*$ case. 
In fact, the HQ expansion of the form factors $g_+$ and $g_-$ gives  
$g_+=\frac{M_1+M_2}{2\sqrt{M_1M_2}}\xi_{IW}(\omega)[1+O(1/m_Q]$ and 
$g_-=-\frac{M_1-M_2}{2\sqrt{M_1M_2}}\xi_{IW}(\omega)[1+O(1/m_Q]$. 
An approximate relation $g_+=-g_-$ can be obtained from this expansion in the limit $M_K^*\ll M_B$ only if the 
generically different combinations $\xi_{IW}(\omega)[1+O(1/m_Q]$ in $g_+$ and $g_-$ evolve in this limit to the same function 
$\xi_{B\to K^*}$ which however goes far from the Isgur-Wise function. 
Let us assume the leading-order IW relations for the form factors with the IW function replaced by the function  
$\xi_{B\to K}$ and $\xi_{B\to K^*}$ for $B\to K$ and $B\to K^*$ transitions, respectively. The 
process-dependent functions $\xi_{B\to K,K^*}$ determined from the GI-OGE QM results for $T_1$ and $F_1$ through 
the relations   
\begin{eqnarray}
\label{xi}
\xi_{B\to K^*}=\frac{4\sqrt{M_B M_K^*}}{M_B+M_K^*}\,T_1,\quad 
\xi_{B\to K}=\frac{2\sqrt{M_B M_K^*}}{M_B+M_K^*}\,F_1
\end{eqnarray}
are shown in Fig. \ref{fig:xi}. The deviations for other form factors of $B\to K^*$ transition found through 
the LO HQS relations with $\xi_{B\to K^*}$ from the lattice-constrained parametrizations can be as much as 20\%. 

Summing up, the dispersion quark model calculates the form factors in the whole 
kinematically accessible decay region. The form factors of the dispersion quark model have the following properties:
they develop a correct HQ expansion in the leading and next-to-leading $1/m_Q$ orders in agreement with 
QCD in heavy-to-heavy transitions provided the soft wave function is concentrated in the region of the
confinement scale; for the case of heavy-to-light transition they 
have the correct scaling properties at small recoil and obey the LO $1/m_Q$ relations between the 
form factors of $V$, $A$, and $T$ currents \cite{iwhl}; and numerically they agree with the lattice results at 
large $q^2$ for the $B\to K^*$ transitions. Thus we expect the dispersion quark model form factors 
to be reliable in the whole kinematically accessible region. 

In the next sections we use the QM form factors evaluated with the GI-OGE wave functions and the 
lattice-constrained parametrizations of Ref. \cite{lat} for analyzing the decay rates and asymmetries 
in rare $B$ decays. 

\begin{table}[hbt]
\caption{\label{table:comparison} Comparison of the results of different approaches on the form 
factors $T_1$, $T_2$, $A_1$ and $A_0$ at $q^2=0$ and $q^2=16.2\;GeV^2$.}
\centering
\begin{tabular}{|c|c|c|c|c|c|c|c|}
Ref. &\multicolumn{2}{c|}{$T_1$}& $T_2$ & \multicolumn{2}{c|}{$A_1$}& \multicolumn{2}{c|}{$A_0$}\\
\hline
\hline
   &$q^2=0$ & $q^2=16.2\;GeV^2$ & $q^2=16.2\;GeV^2$ & $q^2=0$ & $q^2=16.2\;GeV^2$ & $q^2=0$ & $q^2=16.2\;GeV^2$\\
\hline
LCQM \cite{jauswyler} & 0.155  & 0.53& 0.26 &0.26 & 0.45 & 0.32 & $-$ \\
\hline
3ptSR \cite{colangelo} & $0.19\pm0.03$  & $0.5\pm0.05$& $0.13\pm0.03$& $0.37\pm0.03$
& $0.23\pm0.03$& $0.3\pm0.03$& $1.0\pm0.05$ \\
\hline
LCSR \cite{aliev} &0.18& 0.84 & 0.35 & 0.36 & 0.65 & 0.27 & 0.64 \\
\hline 
LCSR+Lat \cite{damir} & $0.15\pm0.04$ & $0.54$ & $0.22$ & $-$  & $-$ & $-$ & $-$  \\
\hline
Lat+ \cite{lat}  
&$0.16^{+0.02}_{-0.01}$ &$0.57\pm0.04$ &$0.21\pm0.04$ &0.29&
$0.44\pm 0.04$ & $0.33$ & $1.28\pm0.07$ \\
\hline 
GI-OGE & 0.177 & 0.53 & 0.248 & 0.33 & 0.44 & 0.44 & 1.20 \\
\hline
ISGW2 &  0.142 & 0.46 & 0.23  & 0.26 & 0.42 & 0.35 & 1.08
\end{tabular}
\end{table} 
 
\newpage
\section{Differential decay rates and lepton asymmetries}
 
In this Section we present formulas for the differential decay rates, forward-backward asymmetries and lepton
polarization asymmetries obtained for both $m_s\ne 0$ and $m_s\ne0$ 
for the transition induced by the effective Hamiltonian Eq. (\ref{heffbtosll}) in the case 
$B \to (K, K^*) (\ell^+ \ell^-)$ and by 
Eq. (\ref{Hneutrino}) in the case $B \to (K, K^*) (\nu\bar{\nu})$. 

Our formulas for the differential decay rates and forward-backward asymmetry coincide with the 
corresponding formulas of Ref. \cite{kruger} and reproduce the formulas of Refs. \cite{gengkao,aliev} 
in the case $m_s=0$ and $m_l\ne 0$ and those of Ref. \cite{colangelo} in the case $m_s=0, m_l=0$. 
For lepton polarization asymmetries our expressions in the limit $m_s=0$ coincide with the results of
\cite{gengkao,aliev}. 

Introducing the dimensionless kinematical variables $\hat{s}
\equiv q^2 / M_B^2$ and $\hat{t} \equiv (P_B - p_{l^+})^2 / M_B^2$, the
double differential decay width for the rare decay $B \to K \ell^+
\ell^-$ can be cast into the form
\begin{eqnarray}
\label{doublewidthBtoP}
{d^2\Gamma(B \to K \ell^+ \ell^-) \over d\hat{s} d\hat{t}} =
{G_F^2 M_B^5 |V_{ts}^* V_{tb}|^2 \alpha_{em}^2 \over 256 \pi^5}
\left [ -\hat{\Pi} \beta_P + 2\hat{m} \delta_{P} \right ],
\end{eqnarray}
where
 \begin{eqnarray}
    \beta_P & = & \left | C_{9V}^{eff}(m_b, q^2) f_+(q^2) + 2(m_b + m_s)
    C_{7\gamma}(m_b) s(q^2) \right |^2 + \left | C_{10A}(m_b) f_+(q^2) \right
    |^2, \nonumber \\
    \hat{\Pi} & = & (\hat{t} - 1) (\hat{t} - \hat{r}) + \hat{s} \hat{t}
    + \hat{m}(1 + \hat{r} + \hat{m} - \hat{s} - 2\hat{t}), \nonumber \\
    \delta_P & = & |C_{10A}|^2 \left \{ \left (1 + \hat{r} - {\hat{s}
    \over 2} \right ) |f_+(q^2)|^2 + (1 - \hat{r}) Re[f_+(q^2)
    f_-^*(q^2)] + {\hat{s} \over 2} |f_-(q^2)|^2 \right \}
 \end{eqnarray}
with $\hat{r} \equiv (M_K / M_B)^2$ and $\hat{m} \equiv (m_{\ell} /
M_B)^2$. After integrating over $\hat{t}$ from $\hat{t}_{min} = [1 +
\hat{r} + 2\hat{m} - \hat{s} - \sqrt{1 - 4\hat{m} / \hat{s}}
\lambda^{1/2}(1, \hat{s}, \hat{r})] /2$ to $\hat{t}_{max} = [1 +
\hat{r} + 2\hat{m} - \hat{s} + \sqrt{1 - 4\hat{m} / \hat{s}}
\lambda^{1/2}(1, \hat{s}, \hat{r})] /2$, where $\lambda(1,\hat s, \hat r)=1+{\hat r}^2+{\hat s}^2-2\hat r - 2\hat
s-2\hat r \hat s$, one obtains the invariant dilepton
mass distribution
\begin{eqnarray}
    \label{widthBtoP}
    {d\Gamma(B \to K \ell^+ \ell^-) \over d\hat{s}} =
    {G_F^2 M_B^5 |V_{ts}^* V_{tb}|^2 \alpha_{em}^2 \over 1536 \pi^5}
    \sqrt{1 - 4\hat{m} / \hat{s}} \lambda^{1/2}(1, \hat{s}, \hat{r})
    \left [ \left (1 + {2\hat{m} \over \hat{s}} \right ) \lambda(1,
    \hat{s}, \hat{r}) \beta_P + 12\hat{m} \delta_P \right ], 
\end{eqnarray}
 
In the case of the decay $B\to K^* \ell^+ \ell^-$ one has
\begin{eqnarray}
    \label{doublewidthBtoV}
    {d^2\Gamma(B \to K^* \ell^+ \ell^-) \over d\hat{s} d\hat{t}} =
    {G_F^2 M_B^{5} |V_{ts}^* V_{tb}|^2 \alpha_{em}^2 \over 512 \pi^5}
    \left [\beta_V^{(1)} + \beta_V^{(2)} + 4\hat{m} \delta_V \right ],
 \end{eqnarray}
with
 \begin{eqnarray}
    \beta_V^{(1)} & = & \left [ (\hat{s} + 2\hat{m} ) \lambda(1,
    \hat{s}, \hat{r}) + 2 \hat{s} \hat{\Pi} \right ] |G(q^2)|^2 + \left
    [ \hat{s} + 2\hat{m} - {\hat{\Pi} \over 2\hat{r}} \right] |F(q^2)|^2
    \nonumber \\
    & & - {\lambda^2(1, \hat{s}, \hat{r}) \over 2\hat{r}} \hat{\Pi}
    |H_+(q^2)|^2 + {\hat{s} - 1 + \hat{r} \over \hat{r}} \hat{\Pi}
    R(q^2), \nonumber \\
    \beta_V^{(2)} & = & 2\hat{s} [2\hat{t} + \hat{s} - \hat{r} - 1 -
    2\hat{m}] R_1(q^2) \nonumber \\
    |G(q^2)|^2 & = & \left | C_{9V}^{eff}(m_b, q^2) M_B g(q^2) - {2C_{7\gamma}(m_b)
    \over \hat{s}} {m_b + m_s \over M_B} g_+(q^2) \right |^2 +
    |C_{10A}(m_b) M_B g(q^2)|^2, \nonumber \\
    |F(q^2)|^2 & = & \left | C_{9V}^{eff}(m_b, q^2) {f(q^2) \over M_B} -
    {2C_{7\gamma}(m_b) \over \hat{s}} {m_b - m_s \over M_B}(1 - \hat{r})
    B_0(q^2) \right |^2 + \left | C_{10A}(m_b) {f(q^2) \over M_B} \right
    |^2, \nonumber \\
    |H_+(q^2)|^2 & = & \left | C_{9V}^{eff}(m_b, q^2) M_B a_+(q^2) -
    {2C_{7\gamma}(m_b) \over \hat{s}} {m_b - m_s \over M_B} B_+(q^2) \right |^2
    + |C_{10A}(m_b) M_B a_+(q^2)|^2, \nonumber \\
    R(q^2) & = & Re \left \{ \left [ C_{9V}^{eff}(m_b, q^2) {f(q^2) \over 
    M_B} - {2C_{7\gamma}(m_b) \over \hat{s}} {m_b - m_s \over M_B} (1 - \hat{r})
    B_0(q^2) \right ] \left [ C_{9V}^{eff}(m_b, q^2) M_B a_+(q^2)\right.\right. \nonumber \\ 
    & &- \left. \left. {2C_{7\gamma}(m_b) \over \hat{s}} {m_b - m_s \over M_B}
    B_+(q^2) \right ]^* \right \} + |C_{10A}(m_b)|^2 Re[a_+(q^2) f^*(q^2)],
    \nonumber \\ 
    R_1(q^2) & = & Re \left \{ \left [ C_{9V}^{eff}(m_b, q^2) M_B g(q^2) -
    {2C_{7\gamma}(m_b) \over \hat{s}} {m_b + m_s \over M_B} g_+(q^2) \right ] 
    \left [ C_{10A}(m_b) {f(q^2) \over M_B} \right ]^* \right \}\nonumber \\
    & & +Re \left \{ \left [ C_{9V}^{eff}(m_b, q^2) {f(q^2) \over M_B} -
    {2C_{7\gamma}(m_b) \over \hat{s}} {m_b - m_s \over M_B} (1 - \hat{r})
    B_0(q^2) \right ] \left [ C_{10A}(m_b) M_B g(q^2) \right ]^* \right
    \}, \nonumber \\
    B_0(q^2) & = & g_+(q^2) + g_-(q^2) {\hat{s} \over 1 - \hat{r}},
    \nonumber \\
    B_+(q^2) & = & -\hat{s} M_B^2 {h(q^2) \over 2} - g_+(q^2),
    \nonumber \\
    \delta_V & = & {|C_{10A}|^2 \over 2} \lambda(1, \hat{s}, \hat{r}) 
    \left \{ - 2 |g(q^2) M_B|^2 - {3 \over \lambda(1, \hat{s}, \hat{r})}
    \left |{f(q^2) \over M_B} \right |^2 + {2(1 + k) - \hat{s} \over
    4\hat{r}} |a_+(q^2) M_B|^2 \right. \nonumber \\
    & & +\left. {\hat{s} \over 4\hat{r}} |a_-(q^2) M_B|^2 + {1 \over
    2\hat{r}} Re[f(q^2) a_+^*(q^2) + f(q^2) a_-^*(q^2)] + {1 - \hat{r}
    \over 2\hat{r}} Re[M_B a_+(q^2) M_B a_-^*(q^2)] \right \}.
 \end{eqnarray}
where now $\hat{r} \equiv (M_{K^*} / M_B)^2$. After integrating over the
variable $\hat{t}$ we find
 \begin{eqnarray}
    \label{widthBtoV}
    {d\Gamma(B \to K^* \ell^+ \ell^-) \over d\hat{s}} =
    {G_F^2 M_B^5 |V_{ts}^* V_{tb}|^2 \alpha_{em}^2 \over 1536 \pi^5}
    \sqrt{1 - 4\hat{m} / \hat{s}} \lambda^{1/2}(1, \hat{s}, \hat{r})
    \left[ \left(1 + {2\hat{m} \over \hat{s}} \right ) \beta_V +
    12\hat{m} \delta_V \right],
 \end{eqnarray}
where
 \begin{eqnarray}
    \beta_V & = & 2 \lambda(1, \hat{s}, \hat{r}) \hat{s} |G(q^2)|^2 +
    \left [ 2\hat{s} + {(1 - \hat{r} - \hat{s})^2 \over 4\hat{r}} \right
    ] |F(q^2)|^2 + {\lambda^2(1, \hat{s}, \hat{r}) \over 4\hat{r}}
    |H_+(q^2)|^2\nonumber \\
    & &-{\lambda(1, \hat{s}, \hat{r}) \over 2\hat{r}} (\hat{s} - 1 +
    \hat{r} ) R(q^2).
 \end{eqnarray}
The effective Hamiltonian for the $B\to(K,K^*)(\nu\bar{\nu})$ transition (\ref{Hneutrino})
may be obtained from the corresponding Hamiltonian for the 
$B \to (K, K^*) (\ell^+ \ell^-)$ transition (\ref{heffbtosll}) 
by the following replacements: 
\begin{equation}
\label{repl}
\hat m\to 0, \; C_{7\gamma}\to0, \; C^{eff}_{9V}\to \frac{X(x_t)}{\sin^2(\theta_W)},\;
C_{10A}\to -\frac{X(x_t)}{\sin^2(\theta_W)}.
\end{equation}
Hence, expressions for the decay rates in $B\to(K,K^*)(\nu\bar{\nu})$ can be obtained from the 
corresponding formulas for $B\to(K,K^*)(\ell^+ \ell^-)$ by the replacement (\ref{repl}). 

For the decays $B \to K^* \ell^+ \ell^-$ a very interesting
quantity is the forward-backward (FB) charge asymmetry
$A_{FB}(\hat{s})$, which is defined as
 \begin{eqnarray}
    \label{Afb}
    A_{FB}(\hat{s}) = {1 \over {d\Gamma(B \to K^* \ell^+ \ell^-)/
    d\hat{s}}} ~ \left [ \int\limits_0^1 d\cos(\theta) ~ {d^2\Gamma(B
    \to K^* \ell^+ \ell^-) \over d\hat{s} d\cos(\theta)} -
    \int\limits_{-1}^0 d\cos(\theta) ~ {d^2\Gamma(B \to K^* \ell^+
    \ell^-) \over d\hat{s} d\cos(\theta)} \right ],
 \end{eqnarray}            
where $\theta$ is the angle between the charged lepton $\ell^+$ and the
$B$-meson directions in the rest frame of the lepton pair. As is
well known, the FB asymmetry is sensitive to the parity structure of
the electroweak interaction. At low values of $q^2$ the
parity-conserving photon exchange is expected to dominate and therefore
the FB asymmetry should be small; on the contary, at large $q^2$ the
contribution of the parity-violating $Z$- and $W$-boson exchanges
becomes relevant, leading to a large asymmetry. Moreover, the FB
asymmetry is sensitive to the relative sign of the Wilson coefficients
\cite{ali} and therefore its measurement could be used as a probe of
the new physics beyond the Standard Model. Explicitly, one has
 \begin{eqnarray}
   \label{Afb_for_BtoV}
    A_{FB}(\hat{s}) = {3\hat{s} \sqrt{1 - 4\hat{m} / \hat{s}}
    \lambda^{1/2}(1, \hat{s}, \hat{r}) R_1(q^2) \over \left (1 +
    {2\hat{m} \over \hat s} \right ) \beta_V + 12\hat{m} \delta_V}.
 \end{eqnarray}

Finally, we will consider also the longitudinal lepton
polarization asymmetry $P_L(\hat{s})$ defined as
\begin{eqnarray}
    \label{lpa} 
    P_L(\hat{s}) = {1 \over {d\Gamma/d\hat{s}}} ~ \left [
    {d\Gamma(h=-1) \over d \hat{s}} - {d\Gamma(h=+1) \over d \hat{s}}
    \right ],
\end{eqnarray}
where $h = +1 (-1)$ means right (left) handed charged lepton $\ell^-$ in
the final state. In case of the rare decay $B \to K \ell^+ \ell^-$ one
has
   \begin{eqnarray}
      \label{plpa}       
      P_L(\hat{s}) & = & {2 \sqrt{1 - 4\hat{m} / \hat{s}} \lambda(1,
      \hat{s}, \hat{r}) \over \left ( 1 + {2\hat{m} \over \hat{s}}
      \right ) \lambda(1, \hat{s}, \hat{r}) \beta_P + 12\hat{m}
      \delta_P} \nonumber \\
      & & Re \left \{ \left [ C_{9V}^{eff}(m_b,q^2) f_+(q^2) + 2 (m_b +
      m_s) C_{7\gamma}(m_b) s(q^2) \right ] C_{10A}^* f_+^*(q^2) \right \},
 \end{eqnarray}
whereas for the process $B \to K^* \ell^+ \ell^-$ one gets
 \begin{eqnarray}
    \label{vlpa}
    P_L(\hat{s}) & = & {2\sqrt{1 - 4\hat{m} / \hat{s}} \over \left ( 1 +
    {2\hat{m} \over \hat{s}} \right ) \beta_V + 12\hat{m} \delta_V} ~ 
    \left [ 2\lambda(1, \hat{s}, \hat{r}) \hat{s} R_G(q^2) + \left (
    2\hat{s} + {(1 - \hat{r} -\hat{s})^2 \over 4\hat{r}} \right )
    R_F(q^2) \right. \nonumber \\
    & & +\left. {\lambda^2(1, \hat{s}, \hat{r}) \over 4\hat{r}}
    R_{H_+}(q^2) - {\lambda(1, \hat{s}, \hat{r}) \over 4\hat{r}} (\hat{s}
    - 1 + \hat{r}) R_R(q^2) \right]
 \end{eqnarray}
where
 \begin{eqnarray}
    R_G(q^2) & = & Re \left \{ \left[ C_{9V}^{eff}(m_b, q^2) M_B g(q^2) -
    {2C_{7\gamma}(m_b) \over \hat{s}} {m_b + m_s \over M_B} g_+(q^2) \right ]
    \left [ C_{10A}(m_b) M_B g(q^2) \right ]^* \right \}, \nonumber \\
    R_F(q^2) & = & Re \left \{ \left [ C_{9V}^{eff}(m_b, q^2) {f(q^2) \over
    M_B} - {2C_{7\gamma}(m_b) \over \hat{s}} {m_b - m_s \over M_B} (1 - \hat{r})
    B_0(q^2) \right ] \left [ C_{10A}(m_b) {f(q^2) \over M_B} \right ]^*
    \right \}, \nonumber \\
    R_{H_+}(q^2) & = & Re \left \{ \left [ C_{9V}^{eff}(m_b, q^2) M_B
    a_+(q^2) - {2C_{7\gamma}(m_b) \over \hat{s}} {m_b - m_s \over M_B} B_+(q^2)
    \right ] \left [ C_{10A}(m_b) M_B a_+(q^2) \right ]^* \right \},
    \nonumber \\ 
    R_R(q^2) & = &  Re \left \{ \left [ C_{9V}^{eff}(m_b, q^2) {f(q^2)
    \over M_B} - {2C_{7\gamma}(m_b) \over \hat{s}} {m_b - m_s \over M_B} (1 - 
    \hat{r}) B_0(q^2) \right ] \left [ C_{10A}(m_b) M_B a_+(q^2) \right
    ]^* \right \} + \nonumber \\
    & & Re \left \{ \left [ C_{9V}^{eff}(m_b, q^2) M_B a_+(q^2) -
    {2C_{7\gamma}(m_b) \over \hat{s}} {m_b - m_s \over M_B} B_+(q^2) \right ]
    \left [ C_{10A}(m_b) {f(q^2) \over M_B} \right ]^* \right \}.
 \end{eqnarray}
Notice that for a left-handed massless neutrino Eq. (\ref{lpa}) yields 
$P_L \equiv-1$. The same result can be obtained also from Eqs. (\ref{repl},\ref{plpa}, \ref{vlpa}).  
\section{Numerical analysis}
In this section we analyze the decay rates and lepton asymmetries in the Standard Model:  
the Wilson coefficients at the scale $\mu\simeq m_b$ are given in Section 2, and 
the $B\to K,K^*$ transition form factors of the dispersion quark model evaluated with the 
GI-OGE model wave functions are adopted. For the $B\to K^*$ transitions we also consider the 
lattice-constrained parametrizations of the form factors of Ref. \cite{lat}. 
We denote the two sets of predictions as QM and Lat, respectively. 
\subsection{Decay rates and dilepton distributions}
First, let us evaluate the $|V_{ts}|$. 
Combining our QM prediction for the $T_2(0)$ with the CLEO data on $B\to K^*\gamma$ \cite{cleo1} we find 
\begin{equation}
|V_{ts}|=0.038\pm 0.005^{\rm exp}. 
\end{equation}
A similar analysis with the lattice-constrained parametrization for $T_2$ \cite{lat} yields 
$|V_{ts}|=0.041\pm 0.005^{\rm th} \pm 0.005^{\rm exp}.$ 

\begin{center}
\begin{figure}
\begin{tabular}{cc}
\mbox{\epsfig{file=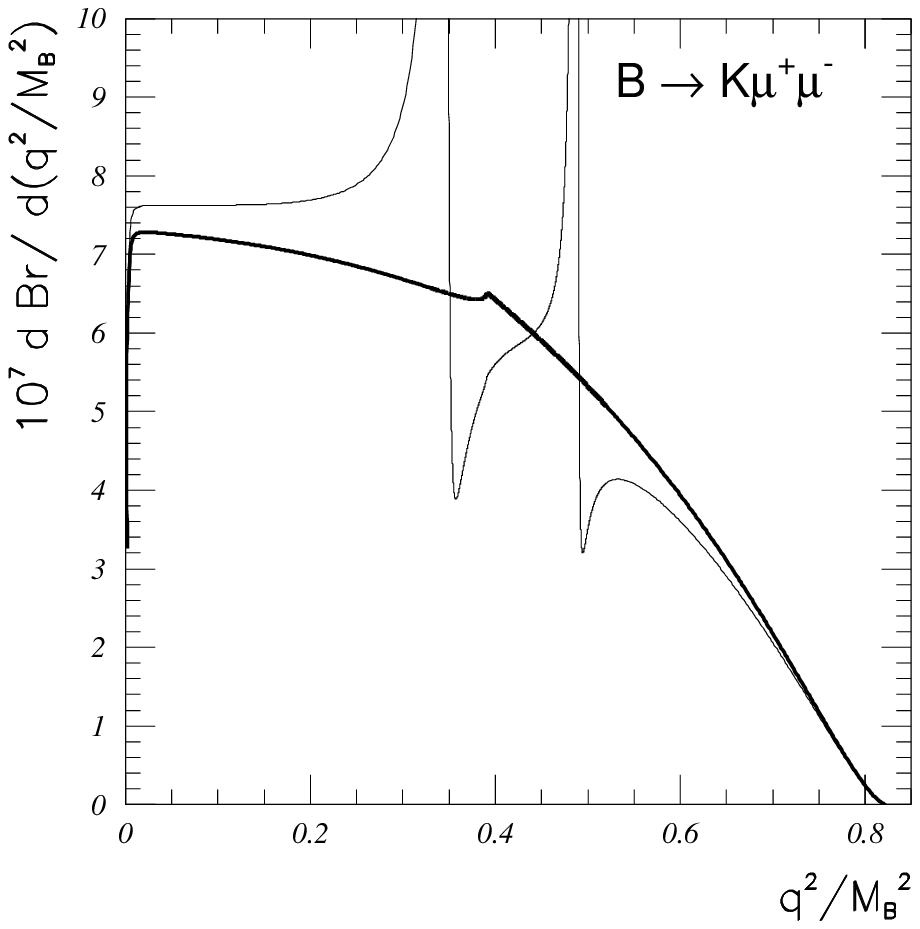,width=7.cm}}
& 
\mbox{\epsfig{file=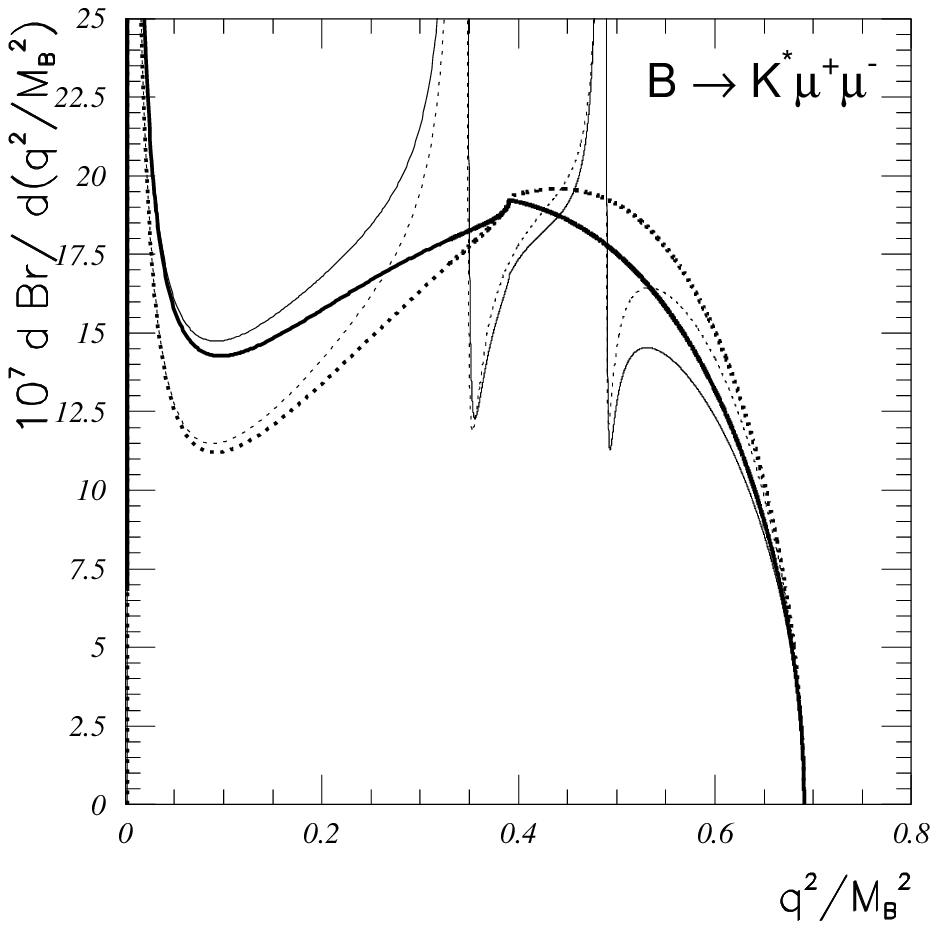,width=7.cm}}\\
\mbox{\epsfig{file=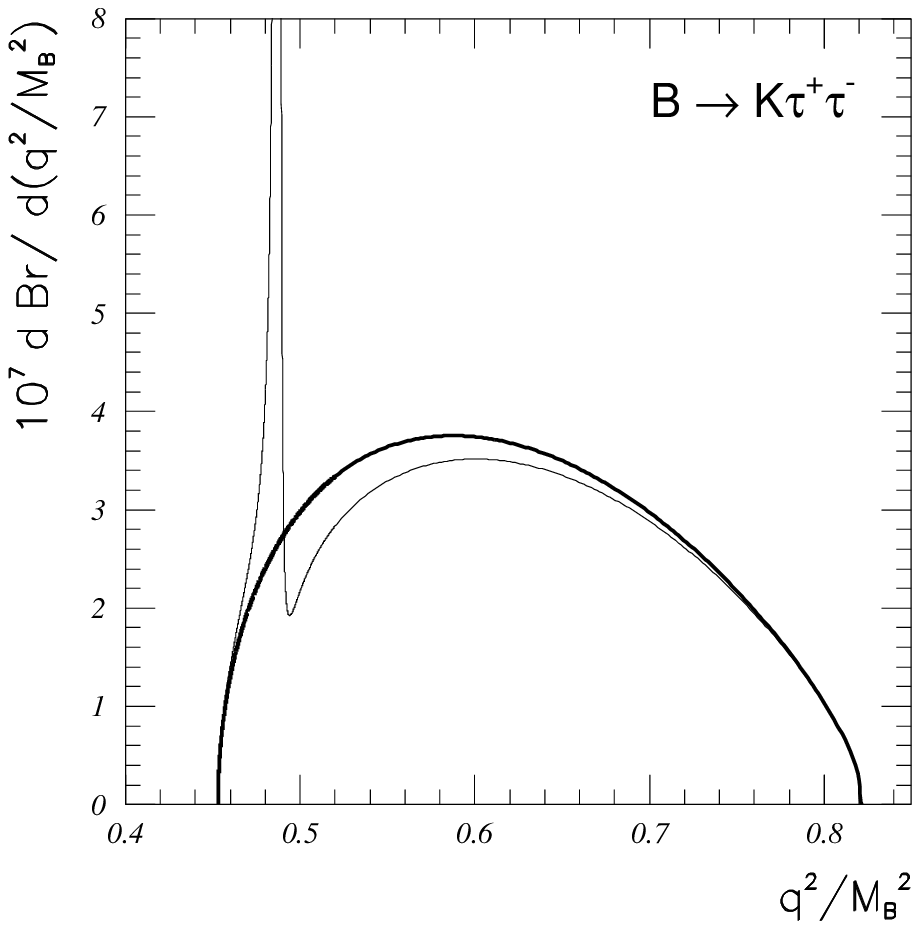,width=7.cm}}
& 
\mbox{\epsfig{file=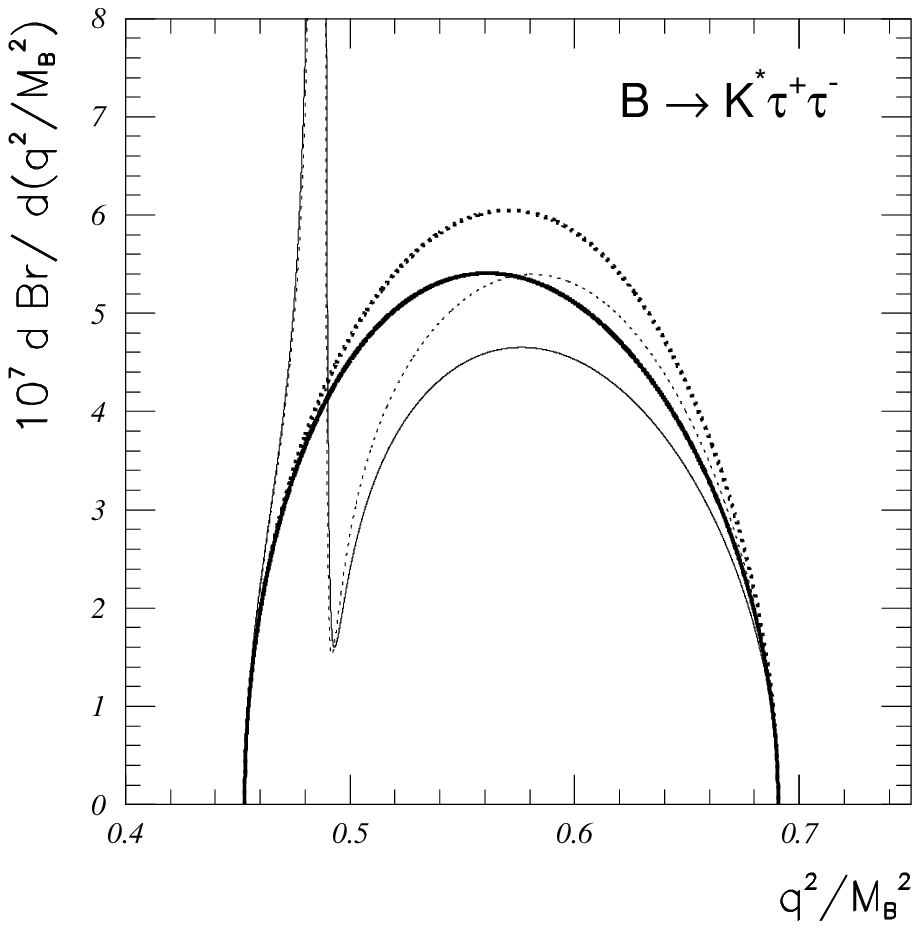,width=7.cm}}
\end{tabular}
\caption{\label{fig:rates}
Differential decay rates $10^7 d\;Br/d(q^2/M^2_B)$ in the decays $B\to K^*\ell^+\ell^-$: 
a. $B\to K\mu^+\mu^-$.    b. $B\to K^*\mu^+\mu^-$.
c. $B\to K\tau^+\tau^-$ , d. $B\to K^*\tau^+\tau^-$.  
Solid - QM form factors (GI-OGE), for $|V_{ts}|=0.038$; dotted - lattice-constrained parametrization of \protect\cite{lat}, 
for $|V_{ts}|=0.041$. Thick lines - nonresonant parts, thin lines - total.}
\end{figure}
\end{center}

The predictions for the dilepton distribution in $B \to (K, K^*) \ell^+ \ell^-$ 
decays are reported in Fig. \ref{fig:rates}, where the non-resonant contributions are also shown separately. 
The total decay rates turn out to be at least one order of magnitude larger
than the non-resonant decay rates. However, the resonant contributions
are strongly peaked in narrow regions around their masses, so that
outside these regions the resonance influence is almost negligible.
This fact allows one to reliably separate the resonant contribution from
the non-resonant one, which is mostly interesting as it contains the information on 
the Wilson coefficients and in principle allows one to measure these coefficients. 
The resonance phases are chosen in accordance with the analysis of Ref. \cite{rescontr}. 

Table \ref{table:rates} summarizes our predictions for the non-resonant branching ratios. 
The branching ratios obtained with the QM and Lat sets of the form factors are given in units $|V_{ts}/0.038|^2$ and 
$|V_{ts}/0.041|^2$, respectively, such that the $B\to K^* \gamma$ branching ratio 
evaluated with $C_{7\gamma}(\mu=5\;GeV)$ in each case is normalized to the 
central CLEO value \cite{cleo1} if this factor is equal to unity. 

Note that the transitions $B \to K^* \mu^+ \mu^-$ and $B \to K^* e^+
e^-$ have different rates, because the amplitude $B \to K^* \ell^+
\ell^-$ has a kinematical pole at $q^2 = 0$, which makes the
corresponding decay rate very sensitive to the lower boundary of the
phase space volume ($q^2 = 4m_{\ell}^2$), while the amplitude $B \to K
\ell^+ \ell^-$ is regular at $q^2 = 0$ and, therefore, insensitive to
the mass of the light lepton. 

One observes a strong sensitivity of the differential decay rates in $B\to K^*$ transitions at low $q^2$: 
the decay rates are results of the interference of various form factors and thus are sensitive to the details of
their $q^2$-behavior. One can see that the Lat and QM form factor sets which both provide a reasonable agreement with
the lattice results at large $q^2$, yield sizeable deviations in the differential distributions at low $q^2$
and hence in the branching ratios. Thus for deriving accurate predictions
for the branching ratios one needs accurate knowledge of the form factors at low $q^2$. 

The decay rates of the $B\to K$ modes turn out to be more stable with respect to variations of the relevant 
form factors (cf. \cite{mns}) and thus might be more perspective for extracting $V_{ts}$ from rare semileptonic 
decays. 
\begin{table}[hbt]
\caption{\label{table:rates}
Non-resonant branching fractions of the radiative and rare semileptonic $B$ decays. 
The branching fractions are evaluated using ${\cal H}_{eff}$ at the scale $\mu=5\;GeV$. Uncertainties connected
with a relevant choice of this low-energy scale $\mu\sim m_b$ are not shown.}
\centering
\begin{tabular}{||l||c|c|c||c||}
Decay      & QM                      & Lat                       & \cite{ali}               &  Exp. \\ 
mode       &$\times |V_{ts}/0.038|^2$& $\times |V_{ts}/0.041|^2$ &$\times |V_{ts}/0.033|^2$ &       \\  
\hline\hline 
$B \to K^* \gamma$ & $4.2\times 10^{-5}$ & $4.2\times 10^{-5}$& $(4.9 \pm 2.0) \times 10^{-5}$ & $(4.2 \pm 1.0) \times 10^{-5}$
\cite{cleo1} 
\\ 
\hline \hline
$B \to K^* e^+ e^-$ & $1.50 \times 10^{-6}$ & $1.45\times 10^{-6}$& $(2.3\pm0.9) \times 10^{-6}$
& $ <1.6 \times 10^{-5}$ \cite{cleo2}\\ 
\hline
$B \to K^* \mu^+ \mu^-$ & $1.15 \times 10^{-6}$& $1.1 \times 10^{-6}$& $(1.5\pm0.6)\times 10^{-6}$
& $< 2.5 \times 10^{-5}$ \cite{cdf}\\ 
\hline
$B\to K^*\tau^+\tau^-$& $1.0\times 10^{-7}$&$1.1\times 10^{-7}$ & $-$ & $-$\\ \hline
$B \to K^* \sum \nu_i \bar{\nu}_i$& $1.5\times 10^{-5}$ & $1.4 \times 10^{-5}$& $(1.1 \pm 0.55)\times 10^{-5}$
& $-$ \\
\hline\hline
$B \to K \ell^+ \ell^-$& $4.4\times10^{-7}$ & $-$&$(4.0\pm1.5)\times 10^{-7}$&$<0.9\times10^{-5}$\cite{cleo2}\\ 
\hline
$B\to K\tau^{+}\tau^{-}$  & $1.0\times 10^{-7}$ & $-$ & $-$ & $-$\\  
\hline
$B\to K\sum\nu_{i}\bar{\nu}_i$ & $5.6\times10^{-6}$ & $-$& $(3.2 \pm 1.6) \times 10^{-6}$& $-$
\end{tabular}
\end{table}

\subsection{Forward-backward asymmetry}

\begin{center}
\begin{figure}
\begin{tabular}{cc}
\mbox{\epsfig{file=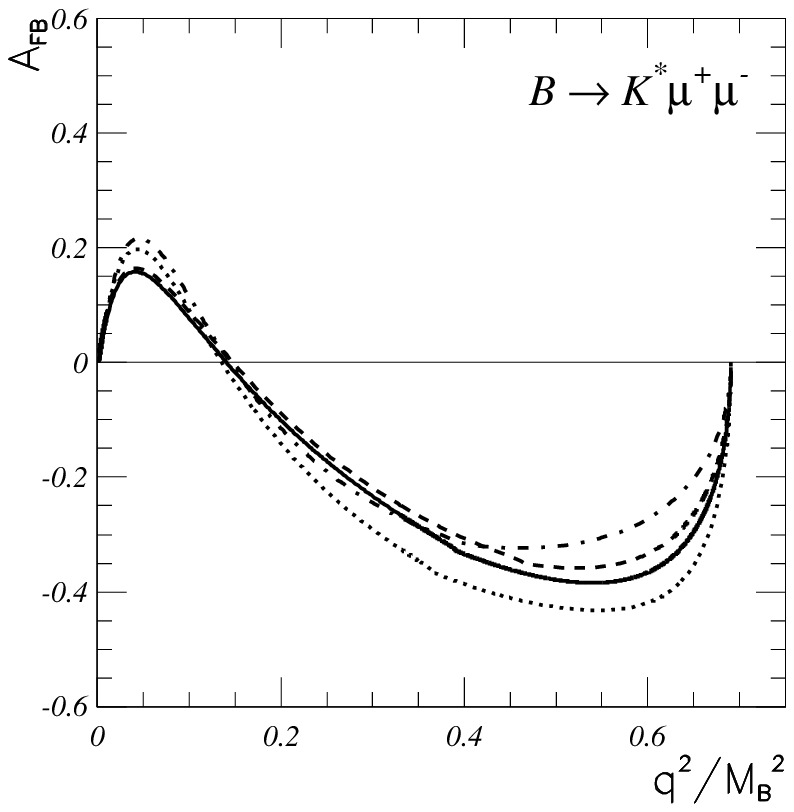,width=7.cm}}
& 
\mbox{\epsfig{file=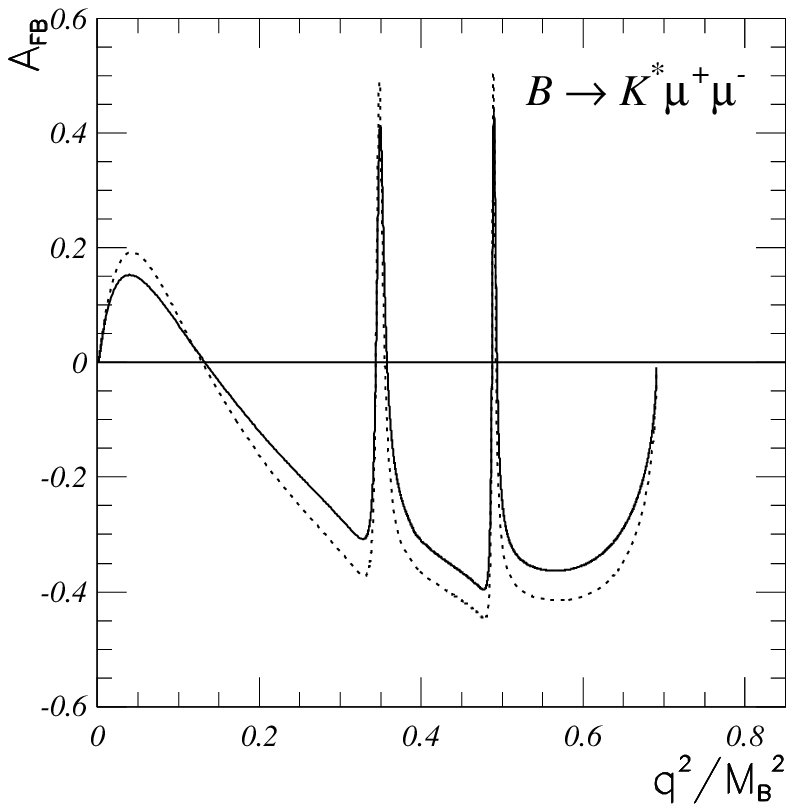,width=7.cm}}\\
\mbox{\epsfig{file=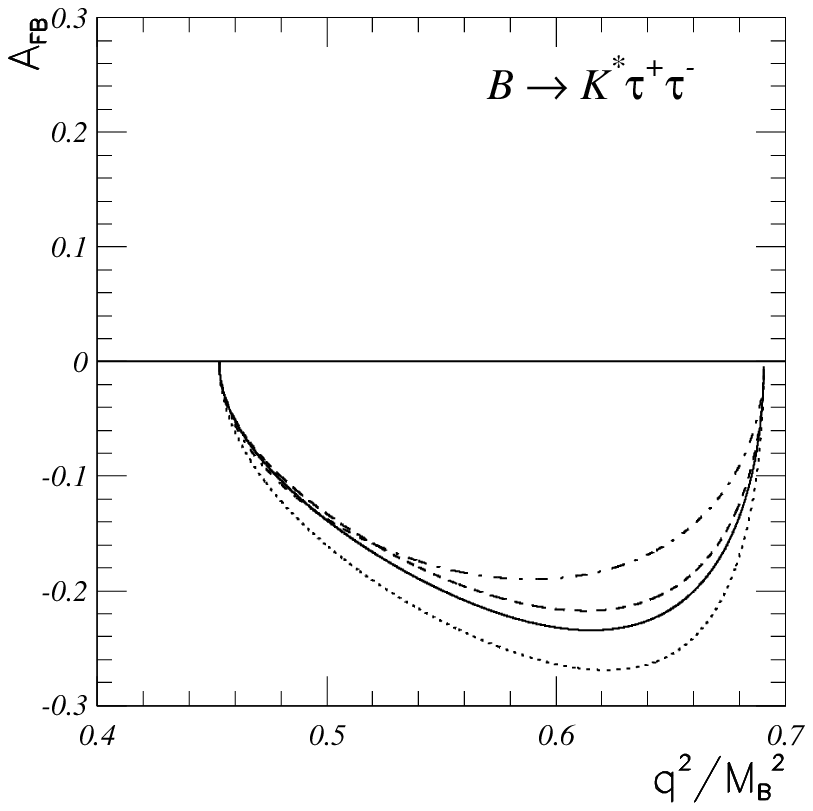,width=7.cm}}
& 
\mbox{\epsfig{file=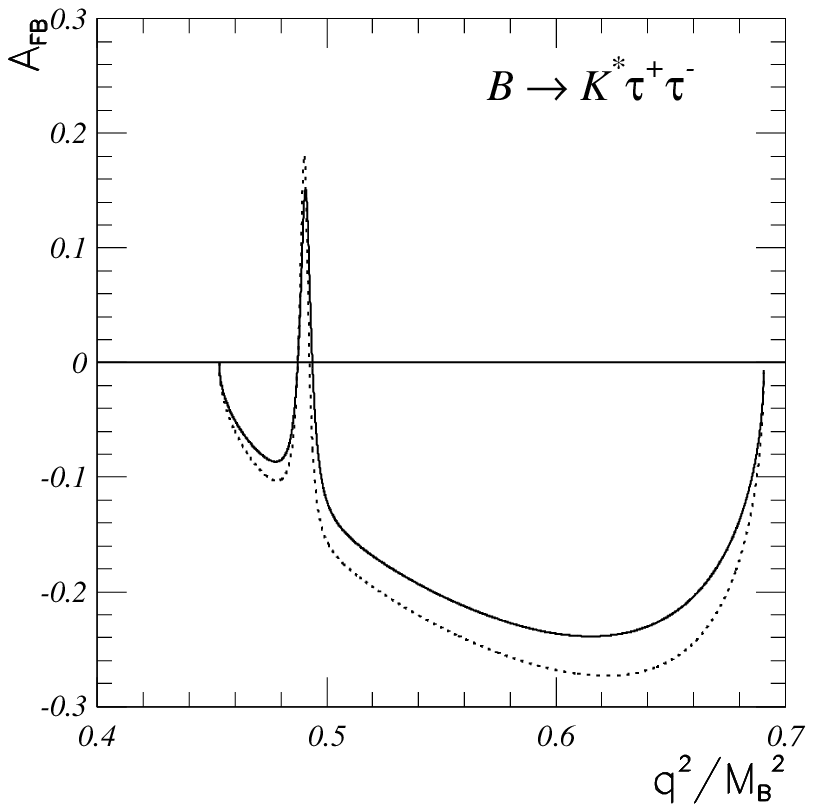,width=7.cm}}
\end{tabular}
\caption{\label{fig:afb}
Forward-backward asymmetries in $B\to K^*\ell^+\ell^-$ transitions. 
a. $B\to K^*\mu^+\mu^-(e^+e^-)$ , non-resonant.  b. The same, total. 
c. $B\to K^*\tau^+\tau^-$ , non-resonant. d. The same, total. 
Solid - GI-OGE, dashed - ISGW2 models, dotted - 
lattice-constrained parametrization of \protect\cite{lat}, dash-dotted -HQS relations. }
\end{figure}
\end{center}

The forward-backward dilepton asymmetries in $B\to K^*\ell^+\ell^-$ decays are presented in Fig. \ref{fig:afb}. 
For comparison, we show also nonresonant $A_{FB}$ evaluated with the form factors for the ISGW2 quark model 
parameters, and obtained assuming the HQS relations between the form factors. 
One can observe a strong sensitivity of the asymmetries to the specific details of the behavior of the relevant
form factors. Notice that the maximum of the asymmetry at $q^2/M_B^2\simeq 0.1$ is mainly proportional to the 
ratio $(V/A_1)^2$. That is why the Lat $A_{FB}$ turns out to be larger in the maximum than the QM asymmetry. 
Nevertheless the general trend of the behavior of $A_{FB}$ for all considered sets of the form factors is 
similar: the nonresonant asymmetry is positive at low $q^2$, has a zero at $q^2/M_B^2\simeq 0.15$, and then 
becomes negative irrespective to the details of the form factor behavior. Let us point out that 
maximum absolute value of the (negative) asymmetry is attained at $q^2/M^2_B\simeq 0.62$ where $|A_{FB}|\simeq 0.4$
both for the QM and Lat sets of the form factors. This is considerably smaller than $|A_{FB}|\simeq 0.6$ reported 
in \cite{colangelo}. This difference is traced back to a very specific behavior of the form factor $T_3$ in 
\cite{colangelo} which contradicts to the results of other approaches and to the approximate HQS relations between
the form factors (see also discussion in \cite{mns}). 

\subsection{Lepton polarization asymmetry} 

\begin{center}
\begin{figure}
\begin{tabular}{cc}
\mbox{\epsfig{file=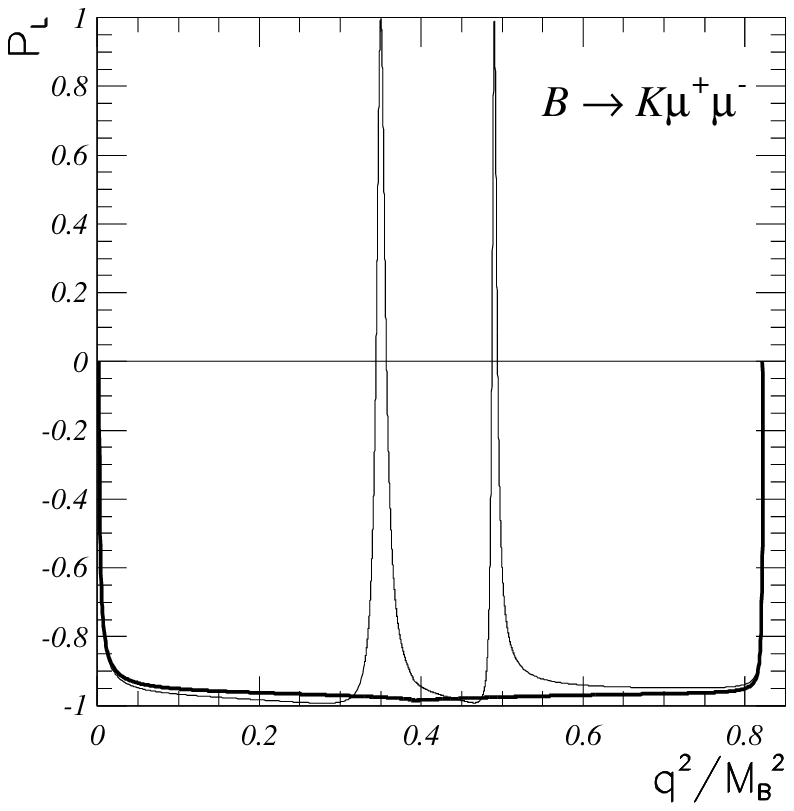,width=7.cm}}
& 
\mbox{\epsfig{file=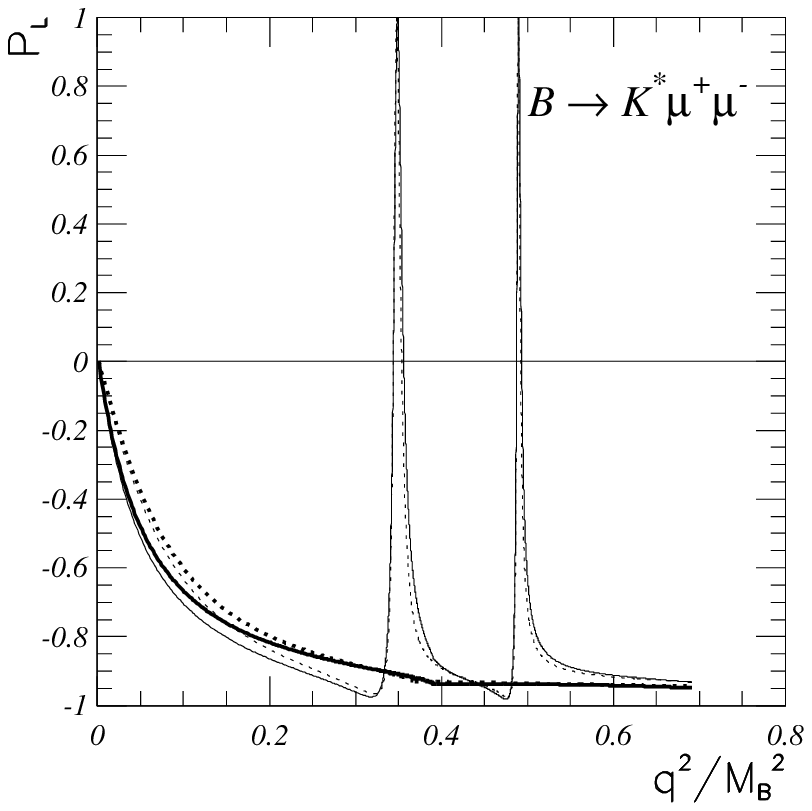,width=7.cm}}\\
\mbox{\epsfig{file=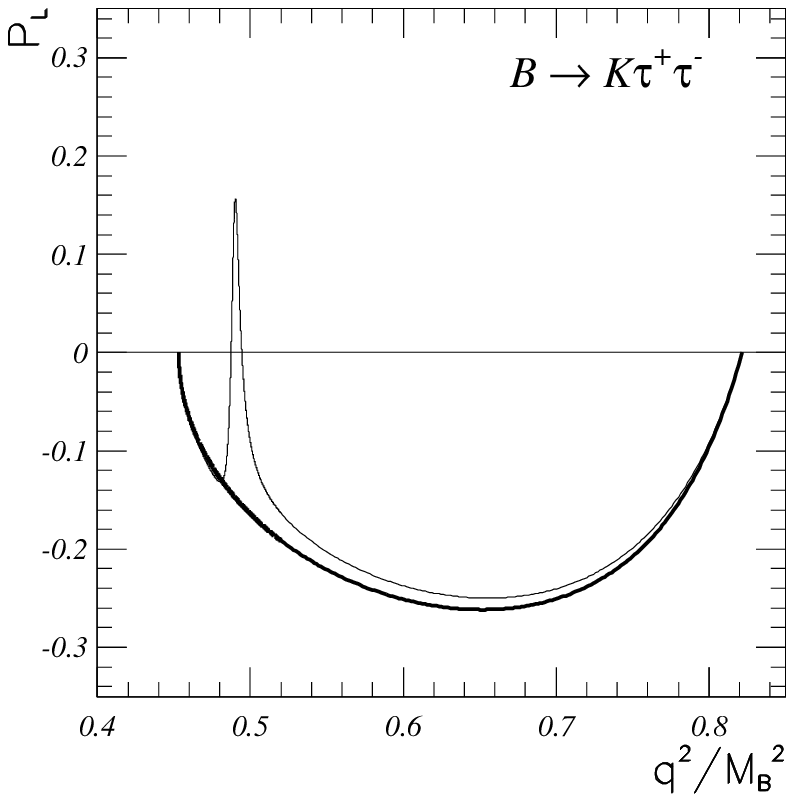,width=7.cm}}
& 
\mbox{\epsfig{file=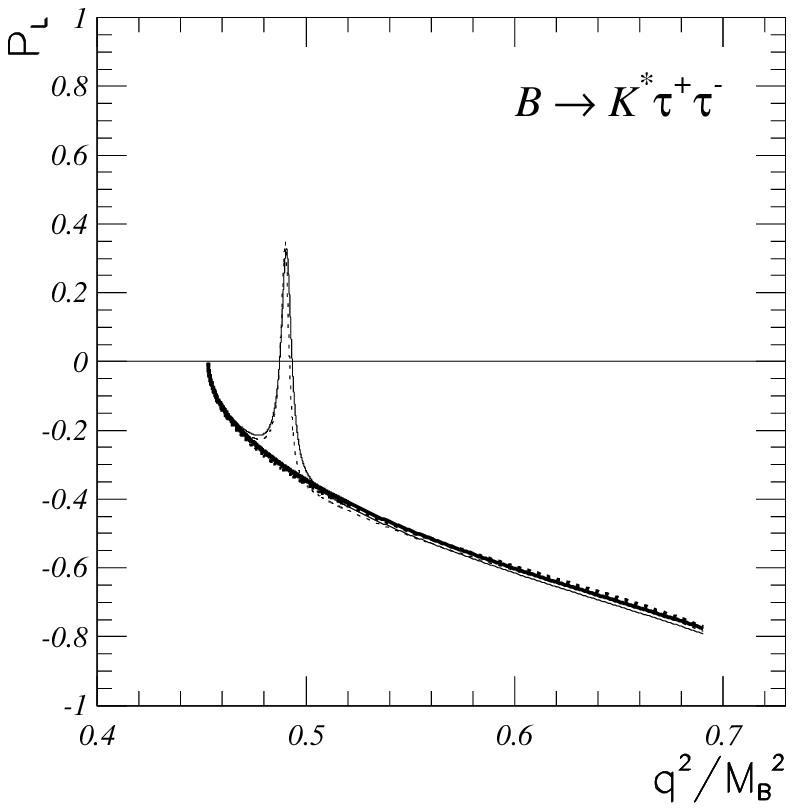,width=7.cm}}
\end{tabular}
\caption{\label{fig:lpa}
Longitudinal lepton polarization asymmetry ($P_L$) in the decays $B\to (K,K^*)\ell^+\ell^-$: 
a. $B\to K\mu^+\mu^-(e^+e^-)$.  b. $B\to K^*\mu^+\mu^-(e^+e^-)$.
c. $B\to K\tau^+\tau^-$. d. $B\to K^*\tau^+\tau^-$.  
Solid - GI-OGE, dotted - lattice-constrained parametrization of \protect\cite{lat}. 
Thick lines - nonresonant, thin lines - total.}
\end{figure}
\end{center}

Fig. \ref{fig:lpa} shows lepton polarization asymmetries $P_L$ for massless and massive leptons. 
For understanding the behavior of $P_L$ it is important to take into account the relationship between
the Wilson coefficients in the SM:
\begin{equation}
\label{wilsonsm}
C_{7\gamma}(m_b)\ll C_{10A}(m_b)\simeq -C_{9V}(m_b).
\end{equation}
In the case of the transition $B\to K\ell^+\ell^-$, $\ell=\mu,e$ a simple analysis of Eq. (\ref{plpa})
yields the following behavior of the nonresonant $P_L$: 
$P_L$ is equal to zero at $q^2=4m_\ell^2$ and $q^2=(M_b-M_K)^2$ due to kinematical reasons, and in
the intermediate region of $q^2$, $P_L$ steeply goes down to the value 
$P_L\simeq 2C_{9V}C_{10A}/(C_{9V}^2+C_{10A}^2)\simeq -1$ independently of the particular behavior of
the $B\to K$ transition form factors. A weak $q^2$-dependence of the nonresonance $P_L$ is due to the
function $h(m_c/m_b, q^2/m_b^2)$ in $C_{9V}^{eff}$. 

In the reaction $B\to K^*\ell^+\ell^-$, $\ell=\mu,e$ the situation is a bit different: 
Now the term in ${\cal H}_{eff}$ proportional to a small $C_{7\gamma}$ contains a photon pole at
$q^2=0$ and thus a parity-conserving photon exchange dominates the decay at low $q^2$ providing a
small value of $P_L$. At large $q^2$ one finds $P_L\simeq-1$ because of just the same reason as in the $B\to K$ 
case with the only difference is that the kinematical zero at $q^2=(M_B-M_{K^*})^2$ is absent. 
In the intermidiate region of $q^2$, the nonresonant $P_L$ is an interplay of the parity-conserving and
parity-violating terms yielding a negative $P_L$ smoothly falling from $0$ to $-1$ in a way largely
independent of the particular $q^2$-dependence of the transition form factors. 

In the total $P_L$ in the reactions $B\to (K,K^*)\ell^+\ell^-$, $\ell=\mu,e$ the $\psi$ and $\psi'$
resonances appear as sharp peaks on a smooth nonresonance background. 
The results of our calculation shown if Fig. \ref{fig:lpa} as well as the results of Ref.
\cite{gengkao,mns} correspond to the picture described above, whereas $P_L$ reported in \cite{aliev} has a different
behavior with $P_L\simeq -0.6$ at large $q^2$ which seems to be very doubtful. 

The lepton polarization asymmetry $P_L$ in the case $B\to (K,K^*)\tau^+\tau^-$ in general follows the
trend of the light leptons case with an important difference: the nonresonant 
$P_L$ does not go down to the value $\simeq-1$ in the kinematically accessible region. 
The $P_L$ again turns out to be largely insensitive to the meson transition form factors. 
In the total $P_L$ one observes only the $\psi'$ peak in the kinematically accessible region. 

\section{CONCLUSIONS}

We have analyzed rare semileptonic transitions $B\to (K, K^*)$ within the 
Standard Model adopting two models for the relevant form factors: a 
relativistic constituent quark model, formulated in a dispersion form, and 
the lattice-constrained parametrization of Ref. \cite{lat}. Our main results 
are as follows:

\begin{itemize}

\item We have presented a dispersion quark model calculation of the 
$B\to K,K^*$ transition form factors in the whole kinematical range of $q^2$. 
Adopting the quark masses and the wave functions of the Godfrey-Isgur 
model \cite{gi} for the hadron spectrum with a switched-off one-gluon 
exchange potential for taking into account only the impact of the 
confinement scale, we have found the resulting form factors to be in 
good agreement with the lattice simulations at large $q^2$. 

The form factors in the dispersion quark model develop the correct expansion 
in the leading and next-to-leading $1/m_Q$ orders for the heavy-to-heavy 
decays, and satisfy the relations between the form factors of the vector, 
axial-vector, and tensor currents valid in the region near the zero-recoil 
point in case of heavy-to-light decays. In addition, the form factors are 
compatible with known analytical constraints. 

Hence, the form factors of the dispersion quark model obey all existing 
rigorous theoretical constraints and agree nicely with the results of 
lattice simulations for the $B\to K^*$ decay at large $q^2$. Moreover, the 
dispersion quark model form factors for the $B\to K^*$ transition agree 
favorably in the whole range of $0<q^2<(M_B-M_K^*)^2$ with a 
lattice-constrained fit \cite{lat} based on the constituent quark picture 
\cite{stech} and an assumption on a single-pole behavior of $A_1(q^2)$. 
Thus we expect to have reliable form factors in the whole kinematically 
accessible decay region. 
\item We have performed a detailed analysis of the non-resonant decay rates 
and asymmetries in $B \to(K,K^*)~(\ell^+\ell^-,\nu\bar{\nu})$ decays in the 
Standard Model and obtained predictions for all exclusive channels using our 
GI-OGE form factors and the lattice-constrained fit to the form factors of 
$B\to K^*$ transition. 
\subitem{i.} Combining our QM result for $T_2(0)$ with the central CLEO 
value for $B\to K^*\gamma$ \cite{cleo1} we estimate the central value 
$|V_{ts}|=0.038$. With the lattice-constrained parametrization of the 
form factors \cite{lat} one finds the central value $|V_{ts}|=0.041$.
\subitem{ii.} The results of the non-resonant branching fractions, 
obtained within the two sets of the form factors, are in good agreement 
if the relevant $|V_{ts}|$ is used in each case. Nevertheless, a better knowledge 
of the relevant form factors around $q^2 = 0$ is still required. 
\subitem{iii.} The differential dilepton distributions in 
$B\to K\ell^+\ell^-$ decays are less sensitive to the details of 
$q^2$-behavior of the form factors than the corresponding distributions 
in $B\to K^*\ell^+\ell^-$ processes. 
Thus, the reaction $B\to K\mu^+\mu^-$ seems to be the most appropriate one 
for the determination of $V_{ts}$ from rare exclusive semileptonic decays. 
\subitem{iv.} The shape of the forward-backward asymmetry in 
$B\to K^*\mu^+\mu^-$ within the SM is almost independent of the 
long-distance contributions: $A_{FB}$ is positive at small $q^2$, has a zero 
at $q^2\simeq 0.15 M^2_B$ and then becomes negative at larger $q^2$. On the 
other hand, the values of $A_{FB}$ in the maximum and the minimum are 
determined by the ratios of the form factors (see also discussion in
\cite{bur}). 
\subitem{v.} The longitudinal lepton polarization asymmetry 
$P_L(B\to K\mu^+\mu^-)$ at all kinematically accessible $q^2$, except 
for the end-points and narrow regions near $\psi$ and $\psi'$, as well as 
$P_L(B\to K^*\mu^+\mu^-)$ at large $q^2$ do not depend on the long-distance 
contributions. In particular, $P_L(B\to K\mu^+\mu^-)\simeq 
2C_{9V}C_{10A}/(C^2_{9V}+C^2_{10A})(\simeq -1$ in the SM), and hence 
$P_L$ directly measures the ratio of the Wilson coefficients $C_{9V}/C_{10A}$ 
at the sacale $\mu\simeq m_b$. Thus, the experimental study of the 
forward-backward asymmetry and the longitudinal lepton polarization 
asymmetry provides an effective test of the Standard Model and its 
possible extentions. 
\end{itemize}
The presented results for the decay rates are essentially based on the 
lattice-constrained constituent quark picture. Further progress in 
obtaining more accurate predictions by combining these approaches may 
be expected on the following way: with increasing the accuracy of the 
lattice predictions one can put forward the determination of the
meson wave functions from the least-$\chi^2$ fit to the lattice results 
at small recoils, basing on our proposed spectral representations 
for the form factors. Then such lattice-constrained quark model would provide 
reliable and accurate form factors at all kinematically accessible $q^2$. 

\section{Acknowledgments} We are grateful to 
D. Becirevic, V. Braun, A. Le Yaouanc, and B. Stech 
for helpful discussions. The work was supported in part by RFBR 
grants 95-02-04808a and 96-02-18121a.

\end{document}